%% file: RectAnt_SM.tex
\documentclass{bmcart}

\usepackage[utf8]{inputenc} 

\usepackage{graphicx}
\usepackage{amsmath}
\usepackage{amssymb}
\usepackage{mathrsfs}
\usepackage{stfloats}
\usepackage{dsfont}
\usepackage{setspace}
\usepackage{array}
\usepackage{bbm}
\usepackage{caption}

\usepackage[T1]{fontenc}
\usepackage{amsmath,amssymb,graphicx,cite}
\usepackage{amsthm,psfrag}
\usepackage{multirow,bigdelim} 
\usepackage{algorithm,algpseudocode}
\usepackage{epstopdf}
\usepackage{color}
\usepackage{booktabs}

\input defs.tex

\usepackage{subcaption}
\startlocaldefs
\endlocaldefs

\begin{document}

\begin{frontmatter}

\begin{fmbox}


\title{Spatial Modulation Based on Reconfigurable Antennas: Performance Evaluation by Using the Prototype of a Reconfigurable Antenna}

\author[
   addressref={aff1},
   email={nvdung1987@gmail.com}
]{\inits{DNV}\fnm{Dung} \snm{Nguyen Viet}}
\author[
   addressref={aff1},                   
   corref={aff1},                       
   email={marco.direnzo@l2s.centralesupelec.fr}
]{\inits{MDR}\fnm{Marco} \snm{Di Renzo}}
\author[
   addressref={aff2},                   
   email={vedaprabhu.b@gmail.com}   
]{\inits{VB}\fnm{Vedaprabhu} \snm{Basavarajappa}}
\author[
   addressref={aff2},                   
   email={bbedia@ttinorte.es}   
]{\inits{BBE}\fnm{Beatriz} \snm{Bedia Exposito}}
\author[
   addressref={aff3},                   
   email={jose.basterrechea@unican.es}   
]{\inits{JB}\fnm{Jose} \snm{Basterrechea}}
\author[
   addressref={aff4},                   
   email={dinhthuy.phanhuy@orange.com}   
]{\inits{DTP}\fnm{Dinh-Thuy} \snm{Phan-Huy}}


\address[id=aff1]{%
  \orgname{Laboratoire des Signaux et Syst\`emes, CNRS, CentraleSupelec, Univ Paris-Sud, Universit\'e Paris-Saclay},
  \street{Plateau du Moulon},
  \postcode{91192},
  \city{Gif-sur-Yvette},
  \cny{France}                          
}
\address[id=aff2]{%
  \orgname{TTI Norte},
  \street{Parque Cientifico y Tecnologico de Cantabria},
  \postcode{Albert Einstein 14, 39011}
  \city{Santander, Cantabria},
  \cny{Spain}
}
\address[id=aff3]{%
  \orgname{University of Cantabria},
  \street{Av. de los Castros, s/n},
  \postcode{39005}
  \city{Santander, Cantabria},
  \cny{Spain}
}
\address[id=aff4]{%
  \orgname{Orange Labs},
  \street{Orange Gardens, 44 avenue de la Republique},
  \postcode{CS 50010, 92326}
  \city{Chatillon Cedex},
  \cny{France}
}



\end{fmbox}


\begin{abstractbox}

\begin{abstract} 
In this paper, we study the performance of spatial modulation based on reconfigurable antennas. Two main contributions are provided. We introduce an analytical framework to compute the error probability, which is shown to be accurate and useful for system optimization. We design and implement the prototype of a reconfigurable antenna that is specifically designed for application to spatial modulation, and that provides multiple radiation patterns that are used to encode the information bits. By using the measured antenna radiation patterns, we show that spatial modulation based on reconfigurable antennas work in practice, and that its performance can be optimized by appropriately optimizing the radiation patterns to use for a given data rate.
\end{abstract}




\end{abstractbox}
%

\end{frontmatter}



%


\section{Introduction} \label{Introduction}
Spatial Modulation (SM) \cite{MDR_Magazine} is a promising low-complexity \cite{Younis_TCOM2013} and energy-efficient \cite{Athanasios_VTC2013} multiple-antenna modulation scheme, which is considered to be especially suitable for application to the Internet of Things (IoT) \cite{OrangeAccess}. For these reasons, SM has attracted the attention of several academic and industrial researchers \cite{MDR_HWU}. A comprehensive description of the main achievements and latest developments on SM research can be found in \cite{MDR_ProcIEEE}-\cite{MDR_IM}. Some pioneering and recent experimental activities can be found in \cite{YounisPractical}-\cite{Liu_2}. The research literature on SM is vast, and various issues have been tackled during the last few years, which include the analysis of the error probability \cite{MDR_TVTMar2012}, the design and analysis of transmit-diversity schemes \cite{MDR_TVT2013}, \cite{MDRVietnam_TCOM2014}, and the analysis and optimization of the achievable rate \cite{MDR_TVT2016}, \cite{MDR_TCOM2018}. A recent comprehensive literature survey on SM and its generalizations is available in \cite{MDR_IM}. \\

Among the many SM schemes that have been proposed in the literature, an implementation that is suitable for IoT applications is SM based on reconfigurable antennas (RectAnt-SM) \cite{MDR_Milcom2017}. In SM, the information bits are encoded onto the indices of the antenna elements of a given antenna-array. In RectAnt-SM, by contrast, the information bit are encoded onto the radiation patterns (RPs) of a single-RF and reconfigurable antenna. This implementation have several advantages, especially for IoT applications \cite{OrangeAccess}. \\

In spite of the potential applications of RectAnt-SM in future wireless networks, to the best of the authors knowledge, no analytical framework for computing the error probability of this emerging transmission technology is available. In the present paper, motivated by these considerations, we introduce an analytical framework that allows us to estimate the performance and to optimize the operation of RectAnt-SM. We prove, in particular, that the diversity order of RectAnt-SM is the same as the diversity order of SM, which coincides with the number of antennas at the receiver. \\

In order to substantiate the practical implementation and performance of RectAnt-SM, we design a single-RF and reconfigurable antenna that provides us with eight different RPs for encoding the information bits at a low-complexity and high energy-efficiency. The proposed antenna is designed, and a prototype is implemented and measured in an anechoic chamber. Based on the manufactured prototype, we employ the measured RPs to evaluate the performance of RectAnt-SM. With the aid of our proposed analytical framework, in particular, we show that the error probability can be improved by appropriately choosing the best RPs, among those available, that minimize the average bit error probability. \\

Together with \cite{OrangeAccess}, the results contained in the present paper constitute the first validation of the performance of RectAnt-SM by using a realistic reconfigurable antenna that is capable of generating multiple RPs with adequate spatial characteristics for modulating information bits. \\

The remainder of the present paper is organized as follows. In Section 2, the system model is introduced. In Section 3, the analytical framework of the error probability is described. In Section 4, the prototype of the single-RF and reconfigurable antenna is presented. In Section 5, numerical results are illustrated, and the performance of RectAnt-SM is analyzed. Finally, Section 6 concludes the paper. \\

\textit{Notation}: We adopt the following notation. Matrices, vectors, and scalars are denoted by boldface uppercase (e.g., $\A$), boldface lowercase (e.g., $\a$), and lowercase respectively (e.g., $a$). The element $(u,v)$ of a matrix $\A$ is denoted by $\A_{u,v}$, and the $u$th entry of a vector $\a$ is denoted by $\a_u$. The transpose, complex conjugate, and complex conjugate transpose of $\A$ are denoted by $\A^T$ , $\A^*$ and $\A^H$ respectively. The absolute value of a complex number $a$ is defined by ${\left| a \right|}$. ${\Exx_a}\left\{  \cdot \right\}$ and ${\mathbb{E}_\A}\left\{  \cdot \right\}$ denote the expectation operator of random variable $a$ and matrix $\A$, respectively. $j = \sqrt { - 1} $ is the imaginary unit. $\Pr \left\{  \cdot  \right\}$ denotes probability. $\binom{\cdot}{\cdot}$ denotes the binomial coefficient. The Q-function is defined as $Q(x) = \left( {1/\sqrt {2\pi } } \right)\int_x^\infty  {\exp } \left( { - \frac{{{u^2}}}{2}} \right){\mkern 1mu} du$. The moment generating function (MGF) of random variable $X$ is defined as ${M_X}\left( s \right) = {\Exx_X}\left\{ {\exp \left( { - sX} \right)} \right\}$. The Gamma function is defined as $\Gamma (z) = \int_0^\infty  {{x^{z - 1}}} {e^{ - x}}\,dx$. The modified Bessel function of order zero is defined as ${I_0}\left(  \cdot  \right)$.

\section{System Model}
In this section, we introduce the signal model, the channel model, and the demodulator.

\subsection{Signal Model}
Let us consider an $N_r\times N_t$ multiple-input-multiple-output (MIMO) system that use a $M$-ary signal constellation diagram. In a conventional SM transmission scheme \cite{MDR_Magazine} the data stream is divided into two blocks, where first block of $\log_2(N_t)$ bits is used to identify the index of  the transmitted antenna and $\log_2(M)$ bits are used to identify a symbol of the signal constellation diagram. By assuming that channel state information (CSI) is known at receiver, the objective of the detector is to jointly estimate the active antenna and the data symbol that is transmitted in order to retrieve the entire transmitted bitstream.

Let us now consider the RecAnt-SM transmission scheme. In this case, we consider that the transmitter is equipped with a reconfigurable antenna that is capable of generating $P$ different RPs. In this case, the RP that is used for transmission is also used to encode the information, in addition to the signal constellation diagram \cite{MDR_Milcom2017}. More precisely, the joint combination of antenna's RP and channel constitute the physical resource that is used to encode the information bits. If $P$ RPs are available, then $\log_2(P)+\log_2(M)$ bits of information can be transmitted.

\begin{figure}[!t]
        \centering
        \begin{subfigure}{0.45\linewidth}
            \centering
            \includegraphics[width=0.9\linewidth]{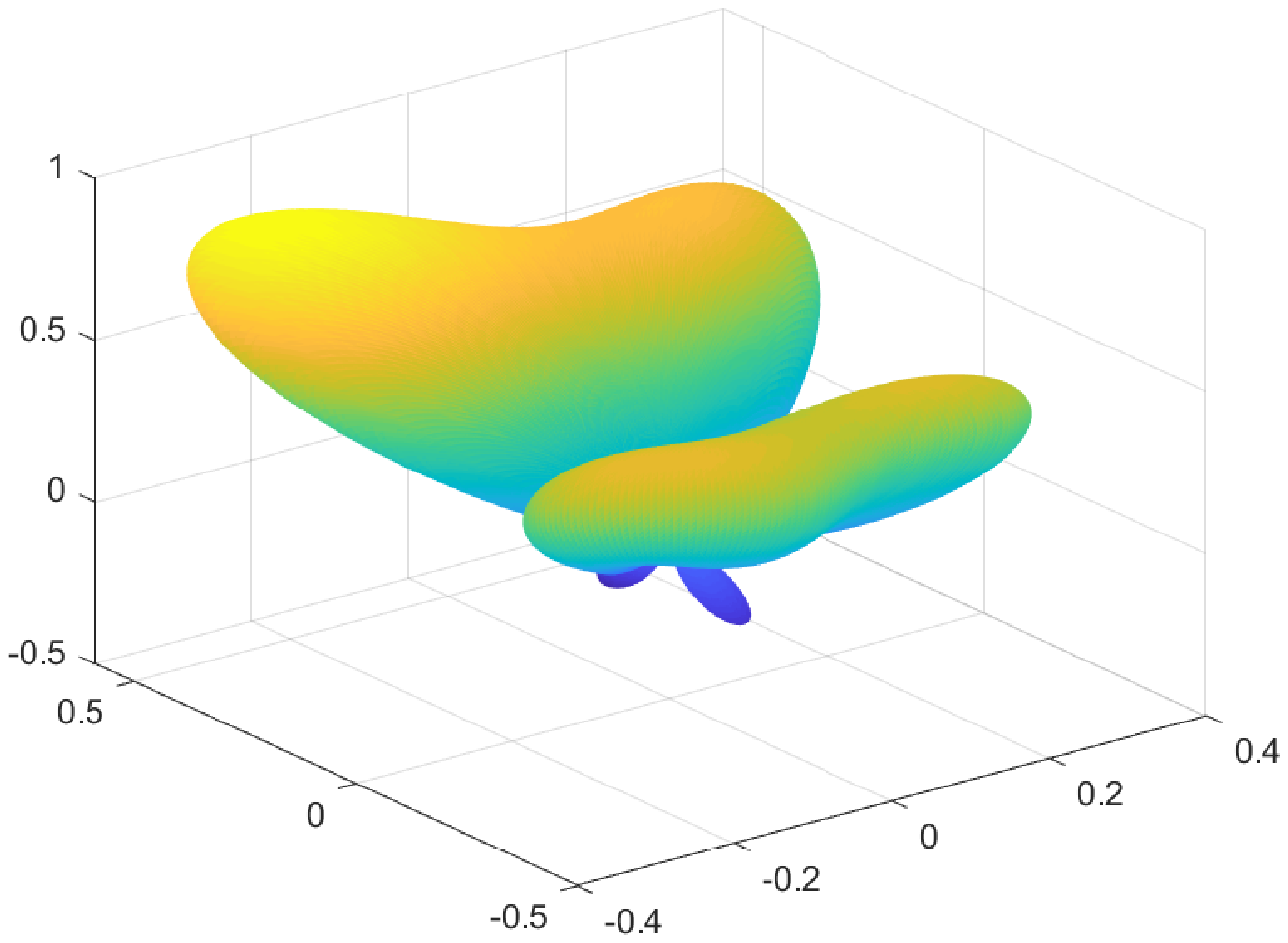}
            \label{RA1}
            \caption{$``00" \to {\text{R}}{{\text{P}}_1 }$}
        \end{subfigure}%
        \begin{subfigure}{0.45\linewidth}
            \centering
            \includegraphics[width=0.9\linewidth]{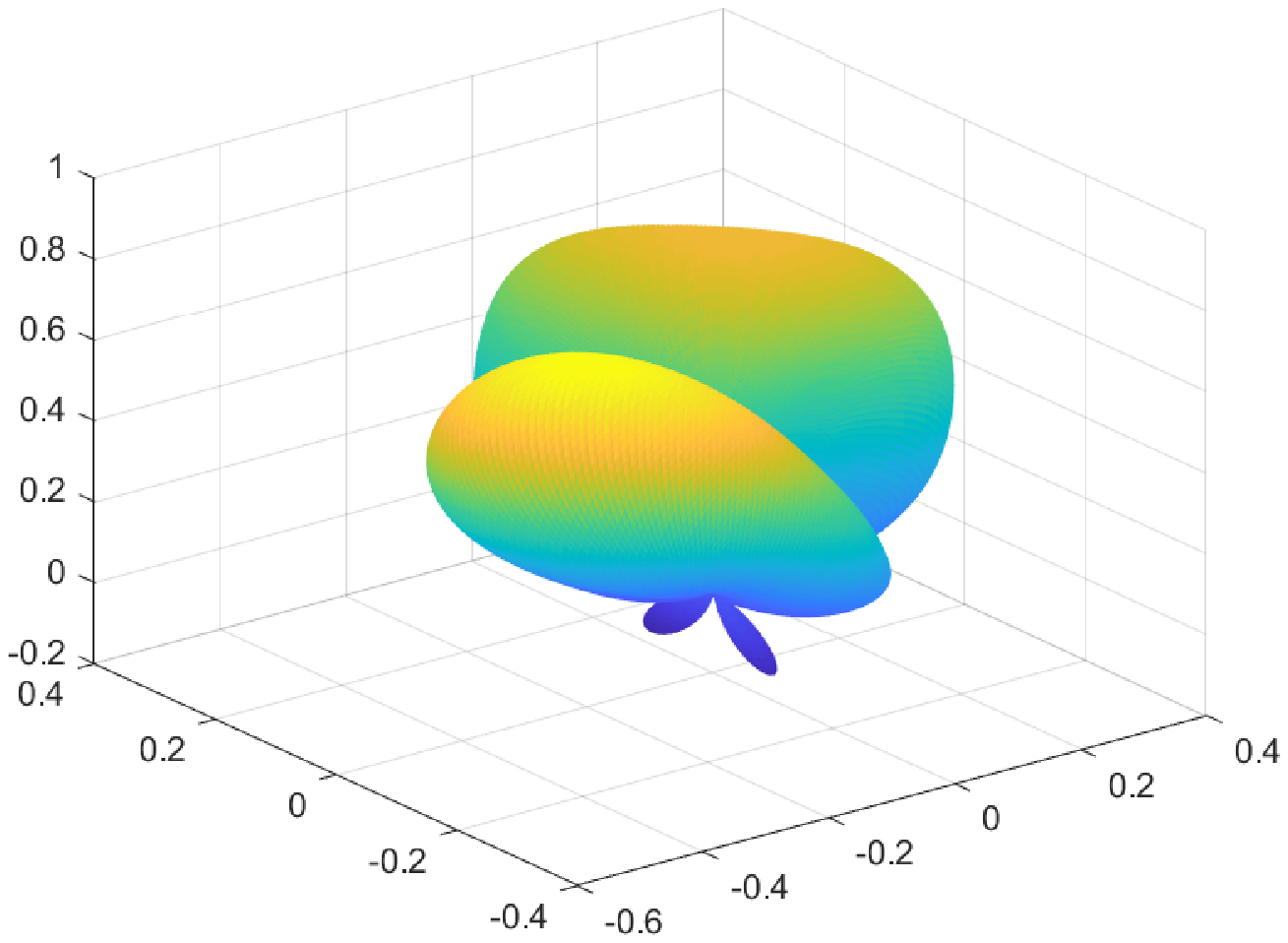}
            \label{RA2}
            \caption{$``01" \to {\text{R}}{{\text{P}}_2}$}
        \end{subfigure}
        \begin{subfigure}{0.45\linewidth}
            \centering
            \includegraphics[width=0.9\linewidth]{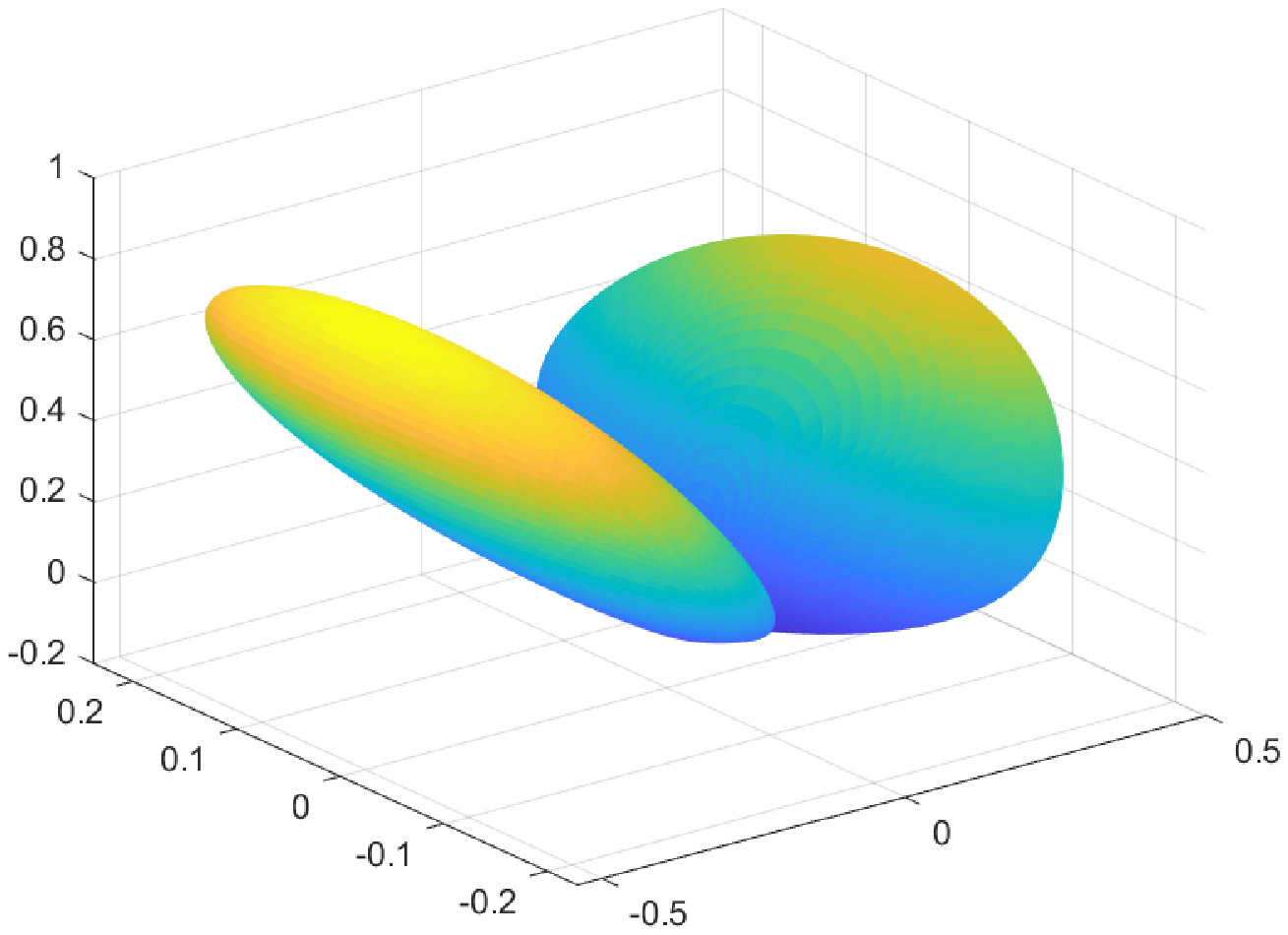}
            \label{RA3}
            \caption{$``10" \to {\text{R}}{{\text{P}}_3}$}
        \end{subfigure}
        \begin{subfigure}{0.45\linewidth}
            \centering
            \includegraphics[width=0.9\linewidth]{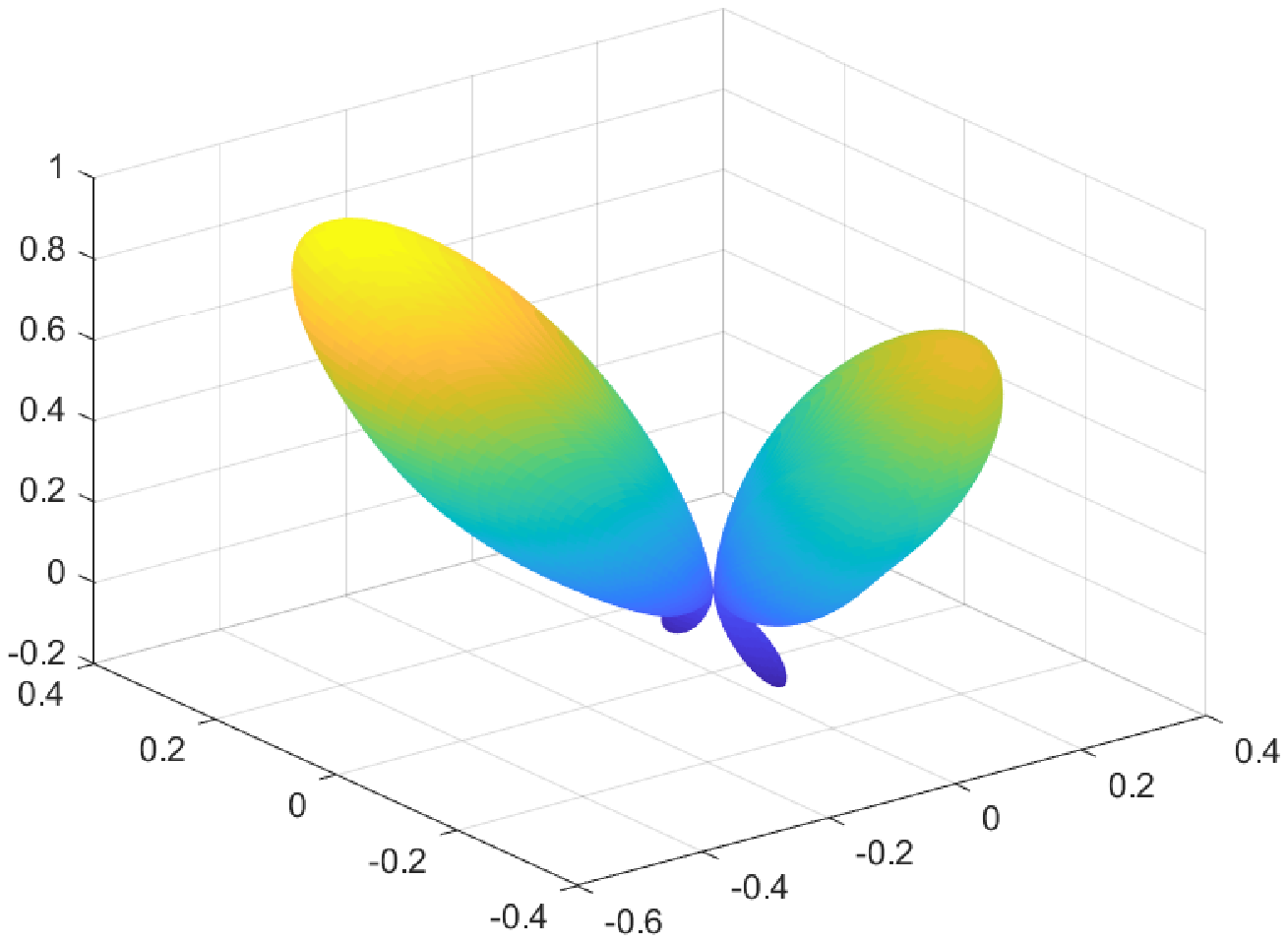}
            \label{RA4}
            \caption{$``11" \to {\text{R}}{{\text{P}}_4}$}
        \end{subfigure}
        \caption{Illustration of RecAnt-SSK with $P = 4$. The RPs are obtained from a designed and manufactured antenna that is described in the sequel.}
        \label{fig:RA}
    \end{figure}

Compared with conventional SM, RecAnt-SM has the advantage of not requiring an array of antennas to enhance the data rate. In addition, RecAnt-SM can be implemented at a low cost, by using simple and compact antennas, where different RPs can be obtained by realizing appropriate circuits that modify the current flowing through the physical antenna. These specific features of RectAnt-SM make it useful for IoT applications. In the sequel, we will discuss an antenna that we have designed and fabricated and that allows us to implement RectAnt-SM in practice. As an example, let us consider $P = M = 4$. Then, $\log_2(P) = 2$ bits are used to identify the RP and $\log_2(M) = 2$ are used to identify a symbol of the signal constellation diagram. Fig. \ref{fig:RA} illustrates a simple example of implementation of RectAnt-SM, by focusing only on the RPs, e.g., RectAnt-SSK (space shift keying).

Let us assume that the $p$th RP and the symbol $x_m$ are selected based on information bits to be transmitted. The $N_r \times 1$ received signal vector can be formulated as follows:
\begin{align} \label{eq:HepSM}
\y &= \sqrt{\rho}\H\e_p x_m + \w
\end{align}
where $\rho$ is the average signal-to-noise-ratio (SNR) at each receive antenna; $\w$ is the $N_r \times 1$ complex additive white Gaussian noise vector of zero mean and unit variance; $\e_p$ for $p = 1, \ldots, P$ is a $P \times 1$ vector whose $p$th entry is equal to one and the other entries are equal to zero; and $\H$ is the $N_r \times P$ channel matrix that accounts for the antenna RPs as well. The vector $\e_p$ allows one to select the specific RP given the bits to be transmitted. The channel matrix $\H$ is introduced in the next section.

RecAnt-SSK constitutes a special case of RecAnt-SM, where the information bits are encoded only into the RPs. In this case, the signal model simplifies as follows:
\begin{align} \label{eq:HepSSK}
\y = \sqrt{\rho}\H\e_p + \n
\end{align}

Based on the signal model in \eqref{eq:HepSM}, the maximum likelihood (ML-) optimum demodulator, by assuming full CSI available at the receiver, can be formulated as follows \cite{Younis_TCOM2013}:
\begin{align}
\left( {\mathord{\buildrel{\lower3pt\hbox{$\scriptscriptstyle\frown$}}
\over q} ,{x_{\mathord{\buildrel{\lower3pt\hbox{$\scriptscriptstyle\frown$}}
\over n} }}} \right) = \mathop {\arg \max }\limits_{{\rm{for}}\:q = 1, \ldots ,P\:{\rm{and}}\:n = 1, \ldots ,M} \left\{ {D\left( {q,{x_n}} \right)} \right\}
\end{align}
where:
\begin{align}
D\left( {q,{x_n}} \right) = \sum\limits_{{n_r} = 1}^{{N_r}} {\left[ {\y_{{n_r}}^*\left(\H_{{n_r},q}{x_n}\right) - \frac{1}{2}{{\left| {{\H_{{n_r},q}}{x_n}} \right|}^2}} \right]}
\end{align}
where $\y_{{n_r}}$ is the $n_r$th entry of $\y$ and $\H_{{n_r},q}$ the entry in the row $n_r$ and column $q$ of $\H$. The demodulator of RectAnt-SSK can be obtained in a similar way.

\begin{figure}[!t]
\centering
\includegraphics[width=0.8\textwidth]{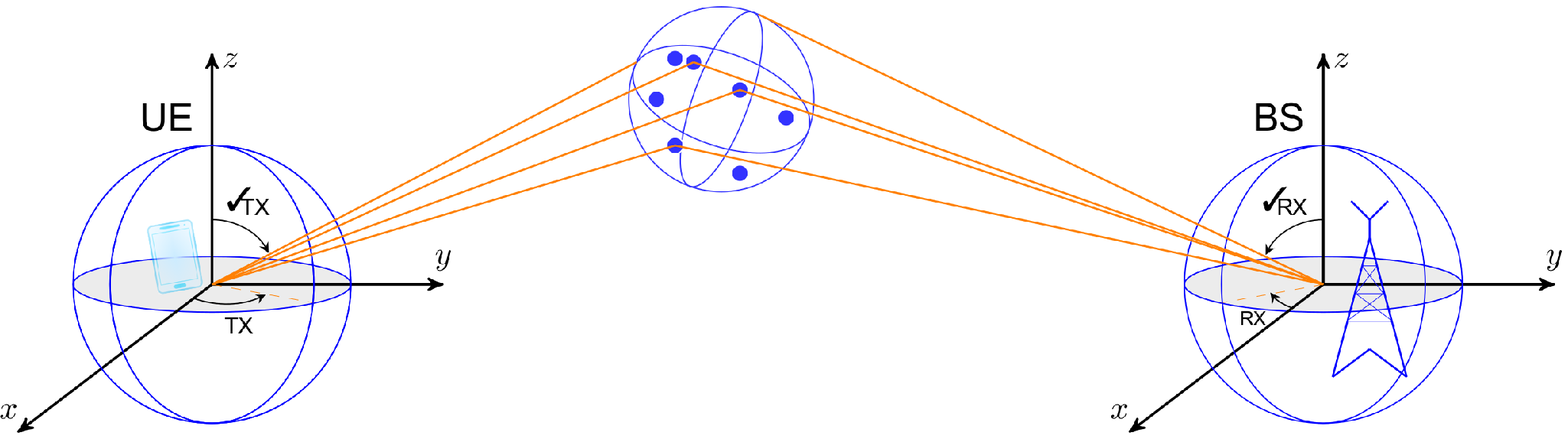}
\caption{Sketched representation of the channel model.}
\label{fig:CM}
\end{figure}
\subsection{Channel Model}
In this section, we introduce the channel model $\H$ that we briefly mentioned in the previous section. As far as RectAnt-SM is concerned, the channel model plays an important role, since the combined effect of RP and channel determines the system performance. Since we consider three-dimensional RPs (see Fig. \ref{fig:RA}), the considered channel model is chosen appropriately. More precisely, the channel model is based on a ray-based and cluster approach similar to \cite{NKDA2015}. A sketched representation of the channel model is given in Fig. \ref{fig:CM}.

For simplicity and without loss of generality, we consider a channel model with a single cluster and with multiple rays. The number of rays is denoted by $K$. The $N_r \times P$ channel matrix, which accounts for the RPs of the reconfigurable antenna as well, can be formulated as follows:
\begin{align}
{{\bf{H}}} = \frac{1}{{\sqrt K }}\sum\limits_{k = 1}^K {{\beta _{k}}} {{\bf{a}}^r}\left( {\theta _{k}^r,\phi _{k}^r} \right){\left( {{{\bf{a}}^t}\left( {\theta _{k}^t,\phi _{k}^t} \right)} \right)^T}
\end{align}
where the normalization factor $1/{\sqrt K }$ preserves the average unit energy of the channel, and the following notation is used:
\begin{itemize}
\item ${\beta _{k}}$ is the fading coefficient of the $k$th ray;
\item $\a^r$ is the $N_r \times 1$ array response vector of the receiver;
\item $\a^t$ is the $N_t \times 1$ array response vector of the transmitter;
\item  $\left( {\theta _{k}^t,\phi _{k}^t} \right)$ are the azimuth and elevation angles of departure (AoD) of the  $k$th ray;
\item $\left( {\theta _{k}^r,\phi _{k}^r} \right)$ are the azimuth and elevation angles of arrival (AoA) of the  $k$th ray.
\end{itemize}

As far as the statistical distributions of $\beta _{k}$, $\theta _{k}$, and $\phi _{k}$ as concerned, Table \ref{tab:Distri} summarizes the most commonly used models. In particular, $\theta_{k}$ is often modeled as a truncated Laplacian random variable, and $\phi_{k}$ is often modeled as a Von-Mises, truncated Gaussian, or an uniform random variable. For each ray, the random variables are assumed to be independent and identically distributed.
\begin{table*}[!t] \footnotesize
\renewcommand{\arraystretch}{1.3}
\caption{Distribution of variables for the considered channel model.}
\label{tab:Distri}
\centering
\begin{tabular}{llll}
\hline
Variable & Distribution & PDF & Range \\ \hline\hline
$\beta_{k}$ & Gaussian & CN(0,1) & $(-\infty,\infty)$ \\ \hline
$\theta_{k}$ & Truncated Laplacian (\cite{ZPPLC2014}) & $f_{\theta}(\theta) = C_L \exp\left(-\frac{\sqrt{2}\mid\theta - \theta_0\mid}{\sigma_L} \right) \sin(\theta)$, \\& & $C_L = \frac{2+\sigma^2_L}{2\sqrt{2}\sigma_L\sin(\theta_0) + 2\sigma^2_L\exp(-\frac{\pi}{\sqrt{2}\sigma_L})\cosh(\frac{\sqrt{2}(\frac{\pi}{2}-\theta_0)}{\sigma_L})}$ & $(0,\pi]$ \\ \hline
\multirow{3}{*}{$\phi_{k}$} & Von-Mises (\cite{ZPPLC2014})& $f_\phi(\phi) = \frac{\exp(\kappa\cos(\phi-\mu))}{2\pi I_0(\kappa)}$ & \multirow{3}{*}{$(-\pi, \pi]$} \\
 & Truncated Gaussian \cite{Pollock2003} & ${f_\phi }(\phi ) = {C_G}\exp \left( { - {{\left( {\frac{{\phi,- {\phi _0}}}{{\sqrt 2 {\sigma _G}}}} \right)}^2}} \right)$ \\& & ${C_G} = \frac{1}{{\sqrt {2\pi } {\sigma _G} {\Phi \left( {\frac{{\pi }}{{{\sigma _G}}}} \right)} }},\Phi (x) = \frac{1}{2}\left( {1 + {\mathop{\rm erf}\nolimits} (x/\sqrt 2 )} \right)$ &  \\
 & Uniform \cite{Pollock2003} & ${f_{\phi} }({\phi} )={\begin{cases}{{\frac{1}{b-a}}}&\mathrm {for} \ a\leq \phi\leq b,\\[8pt]0&\mathrm {otherwise} \end{cases}}$ & \\
 \hline
\end{tabular}
\end{table*}

The specific characteristics of the receiver and transmitter are determined by $\a^r$ and $\a^t$, respectively. As far as the receiver is concerned, we consider, as an example, a uniform linear array with an onmi-directional RP of unit gain. As far as the transmitter is concerned, we consider a reconfigurable antenna with $P$ different RPs. The RPs are denoted as follows:
\begin{align}
{{\cal G}_p}\left( {\theta ,\phi } \right) = \sqrt {{G_p}\left( {\theta ,\phi } \right)} \exp \left( {j{\Omega_p }\left( {\theta ,\phi } \right)} \right)
\end{align}
for $p=1,\ldots,P$, and $G_p(\theta ,\phi )$ and ${\Omega_p}(\theta ,\phi )$ are the amplitude and phase of the $p$th RP, respectively.

Based on these assumptions, $\H_{{n_r},p}$ can be formulated as follows:
\begin{align}
\label{eq:Huv}
{\H}_{n_r,p} = \frac{1}{{\sqrt {K} }} {\sum\limits_{k = 1}^K {{\beta _{k}}} } &\underbrace { \exp (j\parallel {\bf{k}}\parallel {d}(n_r - 1)\sin (\theta _{k}^r)\sin (\phi _{k}^r))}_{{\rm{Receiver\,part}}} \times\nonumber \\
&\underbrace {\sqrt {{G_p}(\theta _{k}^t,\phi _{k}^t)} \exp (j{\Omega _p}(\theta _{k}^t,\phi _{k}^t))}_{{\rm{Transmitter\,part}}}
\end{align}
where $n_r = 1, \ldots ,{N_r}$, $p = 1, \ldots ,P$, $d$ is the distance between adjacent antennas, and $\k(\theta ,\phi )$ is the wavevector defined as follows:
\begin{align}
\resizebox{0.5\textwidth}{!}{${\k}(\theta ,\phi ) = \frac{{2\pi }}{\lambda }{\left[ {\sin (\theta )\cos (\phi ),\sin (\theta )\sin (\phi ),\cos (\theta )} \right]^T}$}
\end{align}
where $\lambda$ is the wavelength.

\section{Average Bit Error Probability}
\label{sec:ABEPPx1}

From \cite{MDR_TVTMar2012}, it is known that the average bit error probability (ABEP) of the system model under analysis can be formulated as follows:
\begin{align}
\label{eq:SNRAPEPPx1Lim}
\ABEP \le \frac{1}{{PM}}\frac{1}{{{{\log }_2}\left( {PM} \right)}}\sum\limits_{p = 1}^P {\sum\limits_{m = 1}^M {\sum\limits_{q = 1}^P {\sum\limits_{n = 1}^M {{N_H}\APEP\left( {\left( {p,{x_m}} \right) \to \left( {q,{x_n}} \right)} \right)} } } }
\end{align}
where APEP denotes the average pairwise error probability, which is the probability of demodulating the RP $q$ and the symbol $x_n$ if the RP $p$ and the symbol $x_m$ have been transmitted, and are the only two possible options, and $N_H$ denotes the number of bits that the latter two constellation points different from each other.

The pairwise error probability (PEP), ${\Pr}_E$, of deciding for the $q$th radiation pattern while the $p$th radiation pattern is transmitted and deciding for the symbol $x_n$ while the symbol $x_m$ is transmitted can be written as follows:
\begin{align}
{\Pr}_E\left( {\left( {p,{x_m}} \right) \to \left( {q,{x_n}} \right)\left| \H \right.} \right) &= \Pr \left( {D\left( {p,{x_m}} \right) < D\left( {q,{x_n}} \right)} \right) \notag\\
&=Q\left( {\sqrt {\frac{\rho }{2}{\gamma _{p,q,{x_m},{x_n}}}\left( \H \right)} } \right)
\end{align}
where:
\begin{align}
{\gamma _{p,q,{x_m},{x_n}}\left( \H \right)} &= \sum\limits_{{n_r} = 1}^{{N_r}} {{{\left| {{\H_{{n_r},q}}{x_n} - {\H_{{n_r},p}}{x_m}} \right|}^2}}
\end{align}

As a result, the APEP can be written as follows:
\begin{align}
\APEP&\left( \left( {p,{x_m}} \right) \to \left( {q,{x_n}} \right) \right) \notag\\
&= {\Exx_{\H}}\left\{ {Q\left( {\sqrt {\frac{\rho }{2}{\gamma _{p,q,{x_m},{x_n}}}\left( \H \right)} } \right)} \right\} \notag\\
&\mathop  = \limits^{(a)}{\Exx_{\H}}\left\{ {\frac{1}{\pi }\int_0^{\pi /2} {\exp} \left( {\frac{{ - \rho{\gamma _{p,q,{x_m},{x_n}}} }}{{4{{\sin }^2}(\vartheta )}}} \right)d\vartheta } \right\}\notag\\
&\mathop  = \limits^{(b)}\frac{1}{\pi }\int_0^{\pi /2} { {{M_{_{{\gamma _{p,q,{x_m},{x_n}}}}}}\left( {\frac{{ - \rho }}{{4{{\sin }^2}(\vartheta )}}} \right)} } d\vartheta
\end{align}
where (a) and (b) follow by applying the Craig's formula \cite{MDR_TVTMar2012} and from the definition of MGF of ${\gamma _{p,q,{x_m},{x_n}}}$, respectively.

\subsection{Setup with $N_r=1$}
\label{subsec:ABEP2x1}
We start by considering the system setup with $N_r = 1$. The APEP is formulated in the following proposition.

\begin{proposition} \label{pro:APEP2x1}
Let $\bar \zeta \left( {K,\vartheta } \right) = \frac{\rho }{{4K{{\sin }^2}(\vartheta )}}$. If $N_r = 1$, the APEP of RecAnt-SM is as follows:
\begin{align}
\label{eq:APEP2x1}
&{\APEP}
= \notag\\ & \frac{1}{\pi }\int_0^{\pi /2} {\int_0^\infty  {\exp } } ( - z){\left( {\int_0^\pi  {\int_{ - \pi }^\pi  {\exp } } \left( { - z\bar \zeta \left( {K,\vartheta } \right)\psi \left( {p,q,{x_m},{x_n},{\theta ^t},{\phi ^t}} \right)} \right){f_\theta }({\theta ^t}){f_\phi }({\phi ^t})\,d{\theta ^t}\,d{\phi ^t}} \right)^K}\,dz\,d\vartheta \\
& \text{where}: \notag\\
&\psi \left( {p,q,{x_m},{x_n},{\theta ^t},{\phi ^t}} \right) = {\left| {\sqrt {{G_q}({\theta ^t},{\phi ^t})} \exp (j{\Omega _q}({\theta ^t},{\phi ^t})){x_n} - \sqrt {{G_p}({\theta ^t},{\phi ^t})} \exp (j{\Omega _p}({\theta ^t},{\phi ^t})){x_m}} \right|^2} \label{eq:Psi}
\end{align}
\end{proposition}
\begin{proof}
See the Appendix.
\end{proof}

In the high-SNR regime, the APEP is given in the following proposition.

\begin{proposition} \label{pro:APEP2x12}
If $N_r = 1$, the APEP of RecAnt-SM in the high-SNR regime is as follows:
\begin{align}
\label{eq:hSNRAPEP2x1}
{\APEP} \le \frac{1}{\rho }{\int_0^\infty  {\left( {\int_0^\pi  {\int_{ - \pi }^\pi  {\exp } } \left( { - z\frac{1}{K}\psi \left( {p,q,{x_m},{x_n},{\theta ^t},{\phi ^t}} \right)} \right){f_\theta }(\theta^t ){f_\phi }(\phi^t )\,d\theta^t \,d\phi^t } \right)} ^K}\,dz
\end{align}
where $\psi$ is defined in (\ref{eq:Psi}).
\end{proposition}
\begin{proof}
See the Appendix.
\end{proof}

Equations (\ref{eq:APEP2x1}) and (\ref{eq:hSNRAPEP2x1}) provide one with accurate performance predictions of the APEP. The analytical expression is, however, quite complex due to the discrete number of rays that are considered in the system model. In the following two propositions, we provide an asymptotic expression of the APEP under the assumption $K \to \infty$. In the sequel, we will show that the simplified analytical expression of the APEP is accurate even for moderate values of $K$.

\begin{proposition} \label{pro:limK}
Assume $K \to \infty$ and $N_r = 1$. The APEP can be simplified as follows:
\begin{align}
\label{eq:SNRAPEP2x1Lim}
&{\APEP} \leq \frac{1}{2}\left( {1 - \sqrt {\frac{{\rho \Theta }}{{\rho \Theta  + 4}}} } \right)
\end{align}
where
\begin{align} \label{eq:Theta}
\Theta  = \int_0^\pi  {\int_{ - \pi }^\pi  {\psi \left( {p,q,{x_m},{x_n},{\theta ^t},{\phi ^t}} \right)} } {f_\theta }({\theta ^t}){f_\phi }({\phi ^t})d{\theta ^t}d{\phi ^t}
\end{align}
where $\psi$ is defined in (\ref{eq:Psi}).
\end{proposition}
\begin{proof}
See the Appendix.
\end{proof}

\begin{proposition} \label{pro:limK1}
Assume $K \to \infty$ and $N_r = 1$. The APEP in the high-SNR regime can be simplified as follows:
\begin{align}
\label{eq:hSNRAPEP2x1Lim}
{\APEP} \le
\frac{1}{\rho\Theta }
\end{align}
where $\Theta$ is defined in (\ref{eq:Theta}).
\end{proposition}
\begin{proof}
See the Appendix.
\end{proof}

By direct inspection of the obtained expression of the APEP, we evince that the error probability is minimized as $\Theta$ increases. As expected, therefore, we evince that the error probability decreases as the difference between the RPs increases. It is worth mentioning, however, that the difference between the RPs is weighted by the distribution of the AoD of the rays. Thus, the APEP depends on both the RPs themselves and the specific characteristics of the channel.

\subsection{Setup with $N_r=2$}
In this section, we study the system setup where two antennas are available at the receiver. The following proposition generalizes Proposition \ref{pro:limK}

\begin{proposition} \label{pro:APEP2x2}
Assume $K \to \infty$ and $N_r = 2$. The APEP in the high-SNR regime can be formulated as follows:
\begin{align}
\label{eq:APEP2x2}
&{\APEP} \le \frac{3}{{{\rho ^2}{{\left( {\int_{ - \pi }^\pi  {\int_0^\pi  {\psi \left( {p,q,{x_m},{x_n},{\theta ^t},{\phi ^t}} \right)} } {f_\theta }({\theta ^t}){f_\phi }({\phi ^t})d{\theta ^t}d{\phi ^t}} \right)}^2}\Psi }}\\
&\begin{array}{c}
\Psi  = 1 - {\left( {\int_{ - \pi }^\pi  {\int_0^\pi  {\cos \left( {\left\| {\bf{k}} \right\|{d}\left( {\sin \left( {{\theta ^r}} \right)\sin \left( {{\phi ^r}} \right)} \right)} \right)} {f_\theta }({\theta ^r}){f_\phi }({\phi ^r})d{\theta ^r}d{\phi ^r}} } \right)^2} - \\
\qquad \:\:\:\:{\left( {\int_{ - \pi }^\pi  {\int_0^\pi  {\sin \left( {\left\| {\bf{k}} \right\|{d}\left( {\sin \left( {{\theta ^r}} \right)\sin \left( {{\phi ^r}} \right)} \right)} \right){f_\theta }({\theta ^r}){f_\phi }({\phi ^r})d{\theta ^r}d{\phi ^r}} } } \right)^2}
\end{array}\notag
\end{align}
where $\psi$ is defined in (\ref{eq:Psi}).
\end{proposition}

By direct inspection of the obtained APEP, we evince that the diversity order is equal to two if two antennas are available at the receiver. This is consistent with conventional SM \cite{MDR_TVTMar2012}.

\subsection{Setup with generic $N_r$}
In this section, we generalize the previous analytical frameworks for $N_r=1$ and $N_r=2$, by considering a generic value of $N_r$.

\begin{theorem} \label{the:Nr}
Assume $K \to \infty$. The APEP in the high-SNR regime can be formulated as follows:
\begin{align}
\label{eq:APEP2xNr}
&{\APEP} \le \frac{{{\alpha _{{N_r}}}}}{{{\rho ^{{N_r}}}{{\left[ {\psi \left( {p,q,{x_m},{x_n},{\theta ^t},{\phi ^t}} \right)} \right]}^{{N_r}}}{\Exx_{{\theta ^r},{\phi ^r}}}\left\{ {F\left( {{\theta ^{1,r}}, \ldots ,{\theta ^{{N_r},r}},{\phi ^{1,r}}, \ldots ,{\phi ^{{N_r},r}}} \right)} \right\}}}\\
&{\alpha _{{N_r}}} = \frac{1}{2}\left( {\begin{array}{*{20}{c}}
{2{N_r}}\\
{{N_r}}
\end{array}} \right) + \sum\limits_{k = 0}^{{N_r} - 1} {{{\left( { - 1} \right)}^{{N_r} - k}}2\left( {\begin{array}{*{20}{c}}
{2{N_r}}\\
k
\end{array}} \right)\frac{{\sin \left( {\pi \left( {{N_r} - k} \right)} \right)}}{{2\pi \left( {{N_r} - k} \right)}}} \notag
\end{align}
where $\psi$ is defined in (\ref{eq:Psi}).
\end{theorem}
\begin{proof}
See the Appendix.
\end{proof}

By direct inspection of the obtained APEP, we observe that the diversity order is equal to $N_r$, as in conventional SM \cite{MDR_TVTMar2012}.

The main limitation of \eqref{eq:APEP2xNr} is that $\Exx_{{\theta ^r},{\phi ^r}}\left\{ {F\left( {{\theta ^{1,r}}, \ldots ,{\theta ^{{N_r},r}},{\phi ^{1,r}}, \ldots ,{\phi ^{{N_r},r}}} \right)} \right\}$ is not analytically tractable and cannot, in general, be formulated in closed-form. In some special cases, however, this is possible. Notably if $N_r = 3$, as reported in the following proposition.

\begin{proposition} \label{pro:2x3}
Assume $K \to \infty$ and $N_r=3$. The APEP in the high-SNR regime can be formulated as follows:
\begin{align} \label{eq:APEP2x3}
{\APEP} \le \frac{{{10}}}{{{\rho ^{{3}}}{{\left[ {\psi \left( {p,q,{x_m},{x_n},{\theta ^t},{\phi ^t}} \right)} \right]}^{{3}}}\Exx_{{\theta ^r},{\phi ^r}}\left\{ {F\left( {{\theta ^{1,r}}, {\theta ^{2,r}}, {\theta ^{{3},r}},{\phi ^{1,r}}, {\phi ^{2,r}} ,{\phi ^{{3},r}}} \right)} \right\}}}
\end{align}
where the following definition holds true:
\begin{align}
&\Exx_{{\theta ^r},{\phi ^r}}\left\{ {F\left( {{\theta ^{1,r}}, {\theta ^{2,r}}, {\theta ^{{3},r}},{\phi ^{1,r}}, {\phi ^{2,r}} ,{\phi ^{{3},r}}} \right)} \right\} = \notag\\
&1 + 2{\left( {{E_1}} \right)^2}{E_3} - 2{\left( {{E_2}} \right)^2}{E_3} + 4{E_1}{E_2}{E_4} - {\left( {{E_3}} \right)^2} - {\left( {{E_4}} \right)^2} - 2{\left( {{E_1}} \right)^2} - 2{\left( {{E_2}} \right)^2}
\end{align}
with
\begin{align}
{E_1} &= {\int_{ - \pi }^\pi  {\int_0^\pi  {\cos \left( {\left\| \k \right\|{d}\left( {\sin \left( {{\theta ^r}} \right)\sin \left( {{\phi ^r}} \right)} \right)} \right)} {f_\theta }\left( {{\theta ^r}} \right){f_\phi }\left( {{\phi ^r}} \right)d{\theta ^r}d{\phi ^r}} }\\
{E_2} &= {\int_{ - \pi }^\pi  {\int_0^\pi  {\sin \left( {\left\| \k \right\|{d}\left( {\sin \left( {{\theta ^r}} \right)\sin \left( {{\phi ^r}} \right)} \right)} \right){f_\theta }\left( {{\theta ^r}} \right){f_\phi }\left( {{\phi ^r}} \right)d{\theta ^r}d{\phi ^r}} } } \\
{E_3} &= {\int_{ - \pi }^\pi  {\int_0^\pi  {\cos \left( {2\left\| \k \right\|{d}\left( {\sin \left( {{\theta ^r}} \right)\sin \left( {{\phi ^r}} \right)} \right)} \right)} {f_\theta }\left( {{\theta ^r}} \right){f_\phi }\left( {{\phi ^r}} \right)d{\theta ^r}d{\phi ^r}} } \\
{E_4} &= {\int_{ - \pi }^\pi  {\int_0^\pi  {\sin \left( {2\left\| \k \right\|{d}\left( {\sin \left( {{\theta ^r}} \right)\sin \left( {{\phi ^r}} \right)} \right)} \right){f_\theta }\left( {{\theta ^r}} \right){f_\phi }\left( {{\phi ^r}} \right)d{\theta ^r}d{\phi ^r}} } }
\end{align}
\end{proposition}
\begin{proof}
See the Appendix.
\end{proof}

\begin{figure}[!t]
\centering
    \includegraphics[width=0.8\linewidth]{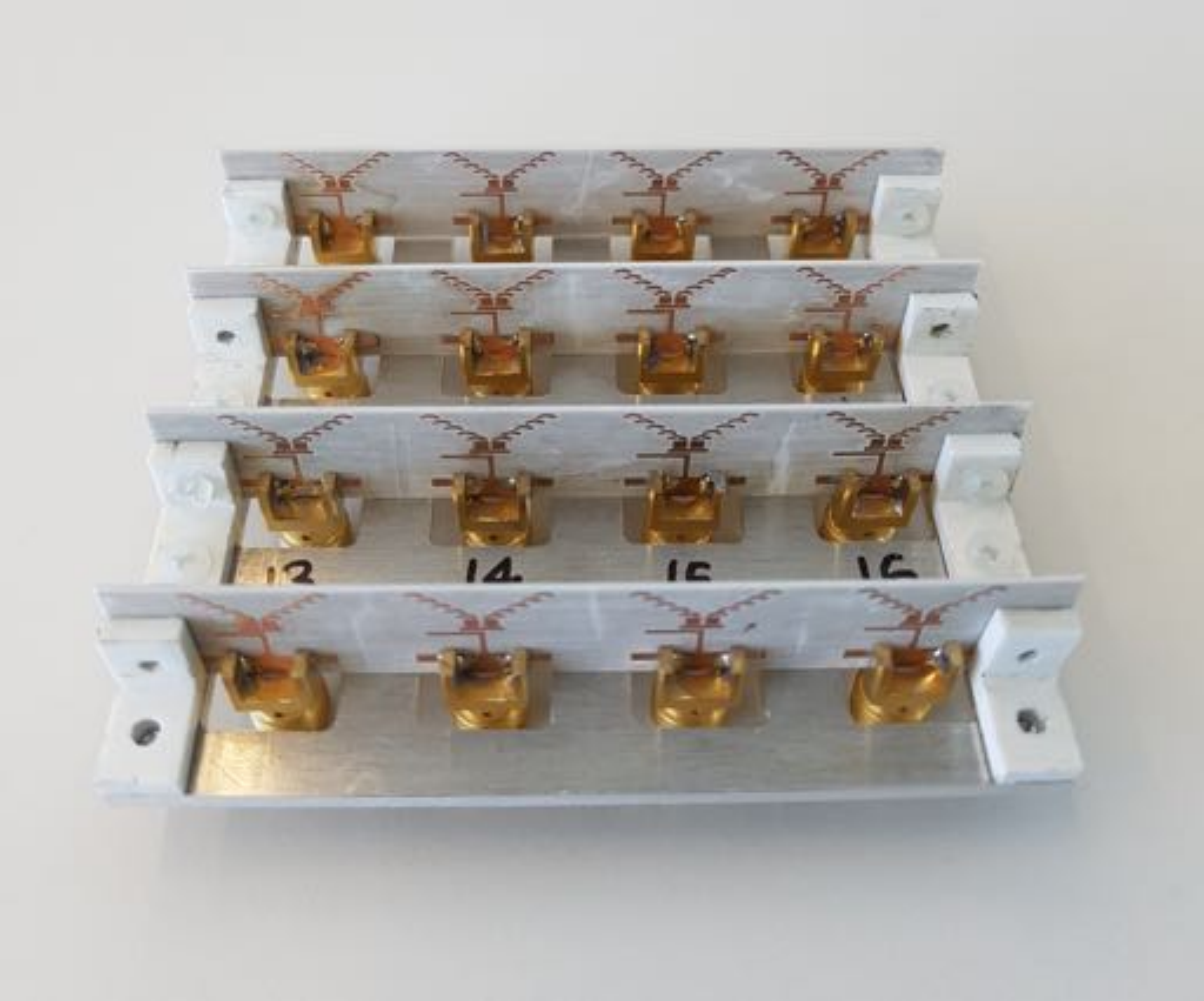}
\caption{Fabricated antenna for implementing RectAnt-SM.}
\label{fig:FAA}
\end{figure}
\begin{figure}[!t]
\centering
\includegraphics[width=0.8\linewidth]{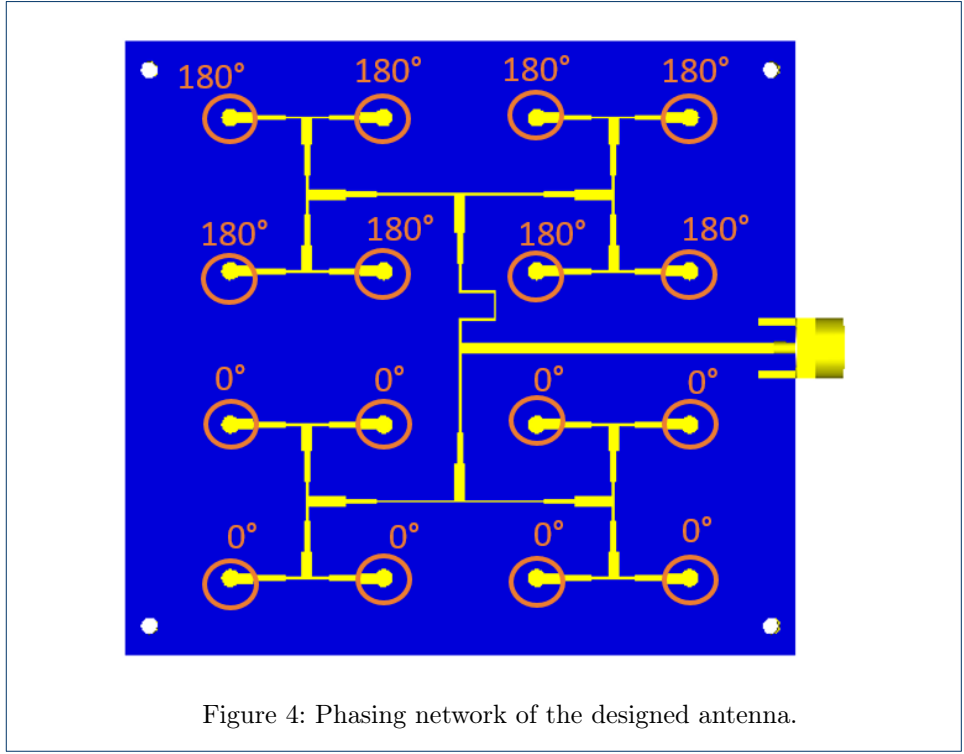}
\caption{Phasing network of the designed antenna.}
\label{fig:PN}
\end{figure}
\section{Antenna Prototype with Reconfigurable Radiation Patterns}
In order to test the performance of RectAnt-SM and to assess its practical feasibility, we have designed and fabricated a reconfigurable antenna that yields multiple radiation patterns. A photo of the antenna prototype specifically designed to implement RectAnt-SM is given in Fig. \ref{fig:FAA}. The proposed antenna needs a single radio frequency chain, and is capable of generating 8 RPs. Four or them are illustrated in Fig. \ref{fig:RA}. The 8 different RPs are obtained by considering a $4 \times 4$ array, as illustrated in Fig. \ref{fig:PN}, and by using a different excitation matrix. Five excitation matrices are reported as follows:
\begin{align}
\mathop {\left[ {\begin{array}{*{20}{c}}
0& + & + &0\\
0& + & + &0\\
0& - & - &0\\
0& - & - &0
\end{array}} \right]}\limits_{\rm{A}} ,\:\mathop {\left[ {\begin{array}{*{20}{c}}
0&0& + & + \\
0&0& + & + \\
 - & - &0&0\\
 - & - &0&0
\end{array}} \right]}\limits_{\rm{B}} ,\:\mathop {\left[ {\begin{array}{*{20}{c}}
0&0&0&0\\
 - & - & + & + \\
 - & - & + & + \\
0&0&0&0
\end{array}} \right]}\limits_{\rm{C}} ,\:\mathop {\left[ {\begin{array}{*{20}{c}}
 - & - &0&0\\
 - & - &0&0\\
0&0& + & + \\
0&0& + & +
\end{array}} \right]}\limits_{\rm{D}} ,\:\mathop {\left[ {\begin{array}{*{20}{c}}
0& - & - &0\\
0& - & - &0\\
0& + & + &0\\
0& + & + &0
\end{array}} \right]}\limits_{\rm{E}}
\end{align}

Each entry of the $4 \times 4$ matrix represents the excitation that is fed into the corresponding antenna of the array. The excitation has unit amplitude and its phase is either $0$ or $180$, which is denoted by the sign ``+'' and ``-'', respectively, in the excitation matrices.

\begin{figure}[!t]
\centering
    \includegraphics[width=0.8\linewidth]{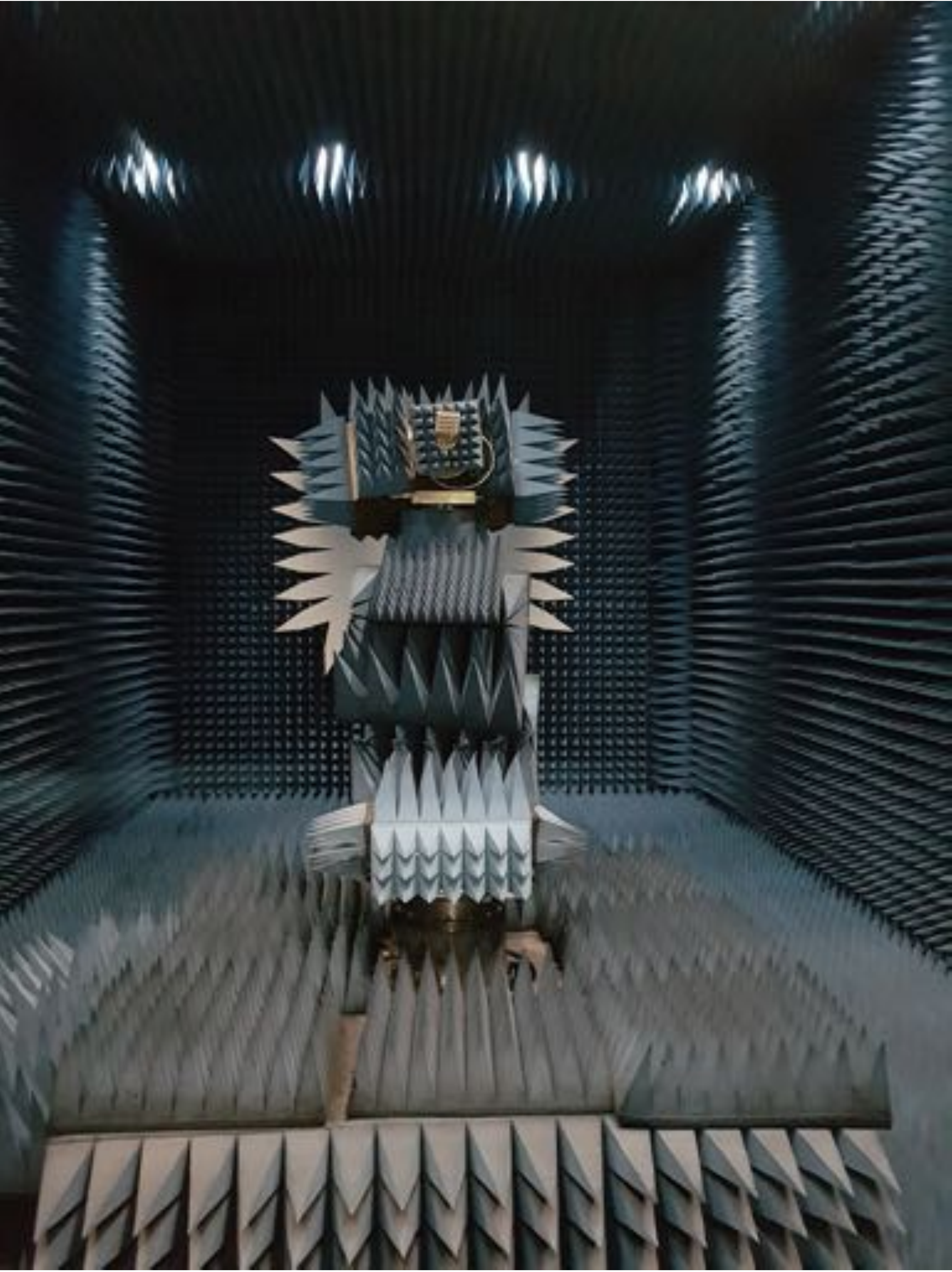}
\caption{Measurement of the antenna prototype in an anechoic chamber.}
\label{fig:AnCham}
\end{figure}
\begin{figure}[!t]
\centering
    \includegraphics[width=0.8\linewidth]{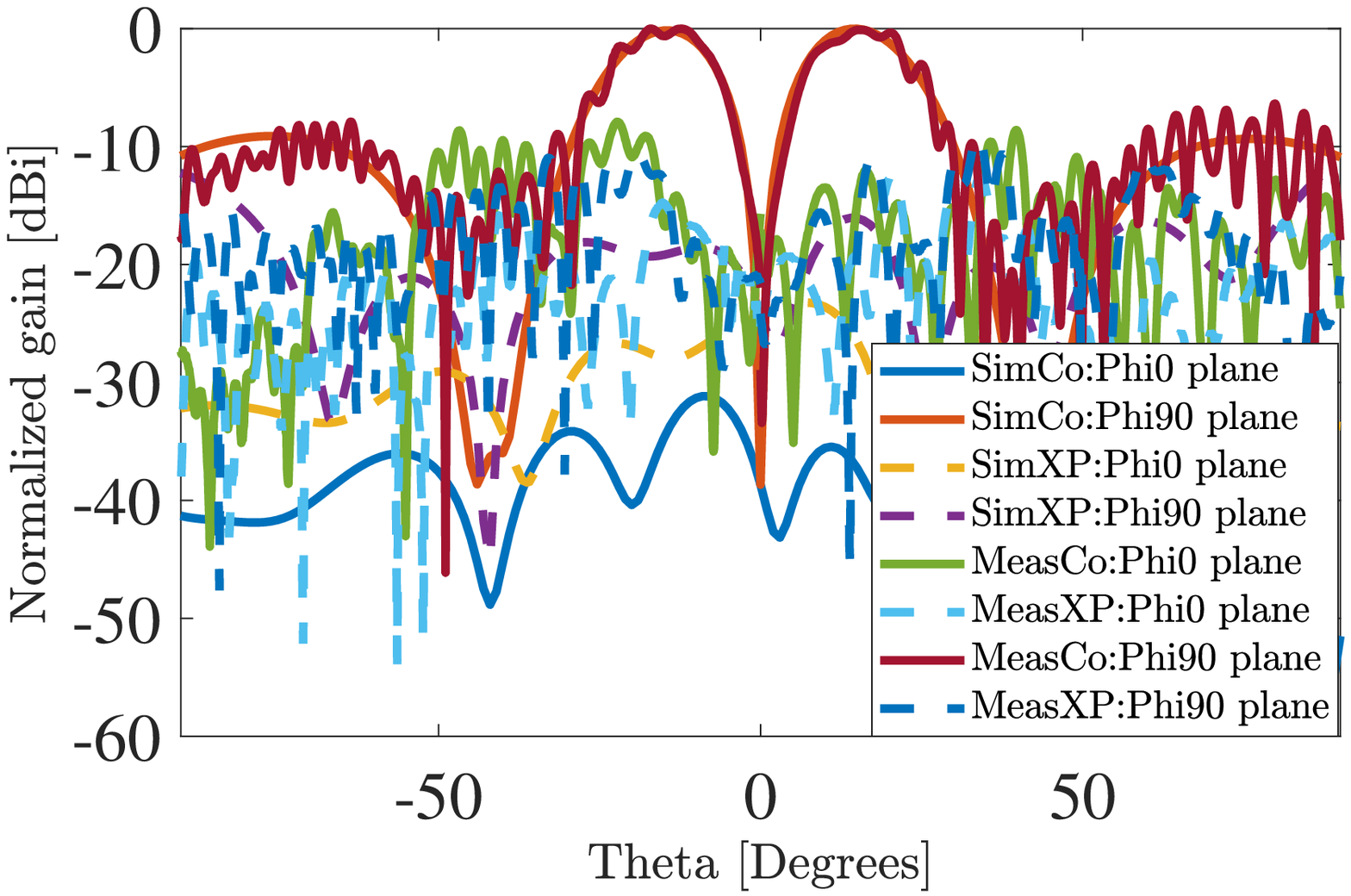}
\caption{Comparison of simulated (Sim) and measured (Meas) RP $\text{RP}_1$. The beam cuts are measured at differently oriented angles (0 and 90 degrees) in space. ``Co'' stands for ``Copolar gain'' and ``XP'' stands for ``Cross Polar gain''.}
\label{fig:RP1SM}
\end{figure}
The designed prototype has been studied and optimized by simulating the $4 \times 4$ antenna array with the full-wave solver Ansys HFSS. Based on the optimized design, a prototype antenna has been fabricated and its RPs have been measured in an anechoic chamber, as illustrated in Fig. \ref{fig:AnCham}. In Fig. \ref{fig:RP1SM}, we report, as an example, the simulated and measured RP $\text{RP}_1$ that is reported in Fig. \ref{fig:RA} as well.

\section{Numerical Results and Discussion}
In this section, we provide some numerical results in order to validate the analytical derivation of the bit error probability, and in order to asses the achievable performance by using the RPs that are obtained from the fabricated antenna prototype. We show, in particular, that by appropriately selecting the RPs that minimize the analytical framework of the error probability, the error probability can be greatly decreased.

\begin{figure}[!t]
\centering
    \includegraphics[width=0.8\linewidth]{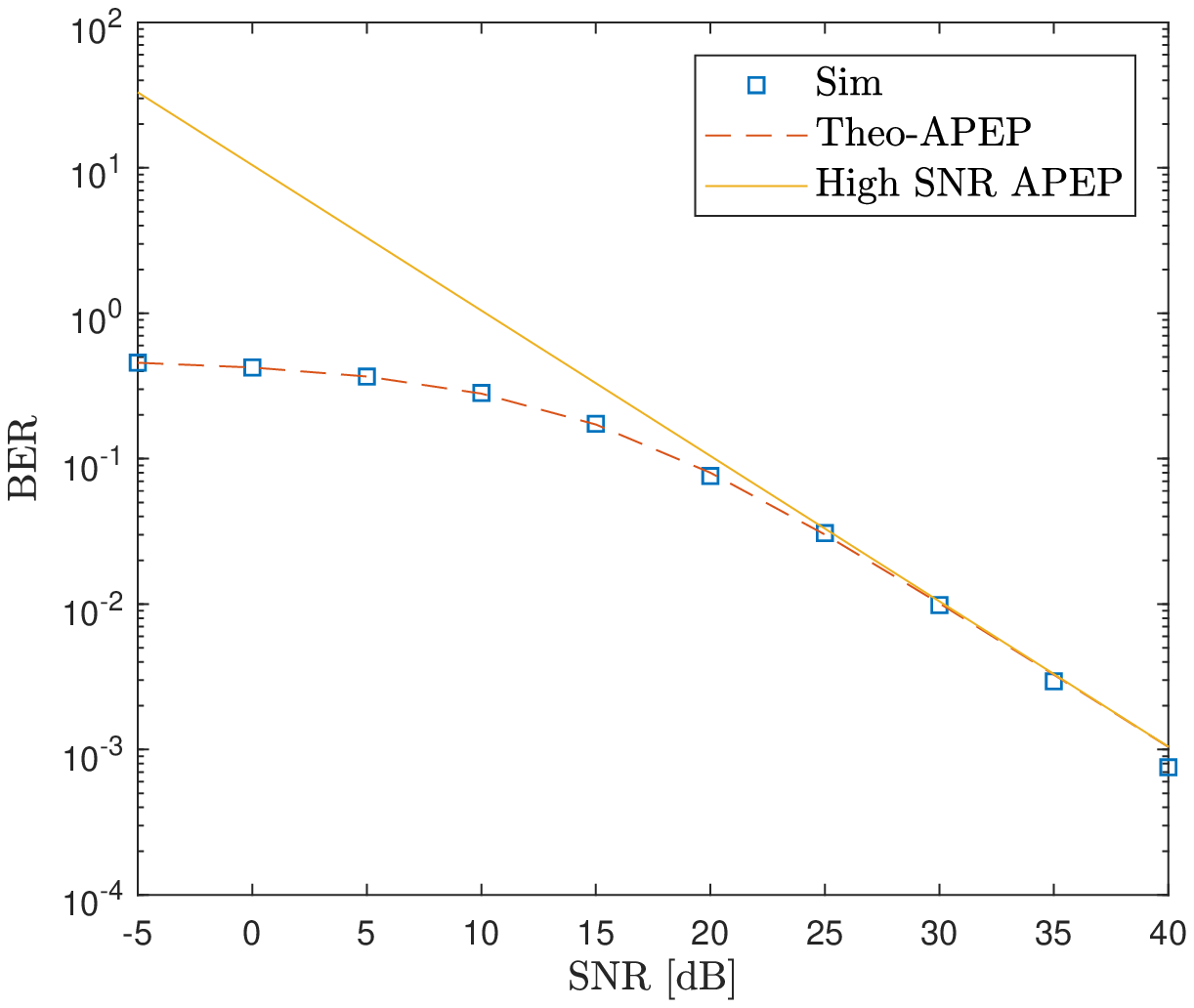}
\caption{ABEP of RecAnt-SM ($P = 2$, $N_r = 1$, $M = 2$ (BPSK)). The results are obtained by using $\text{RP}_1$ and $\text{RP}_2$.}
\label{fig:exactABEP}
\end{figure}
\begin{figure}[!t]
\centering
    \includegraphics[width=0.8\linewidth]{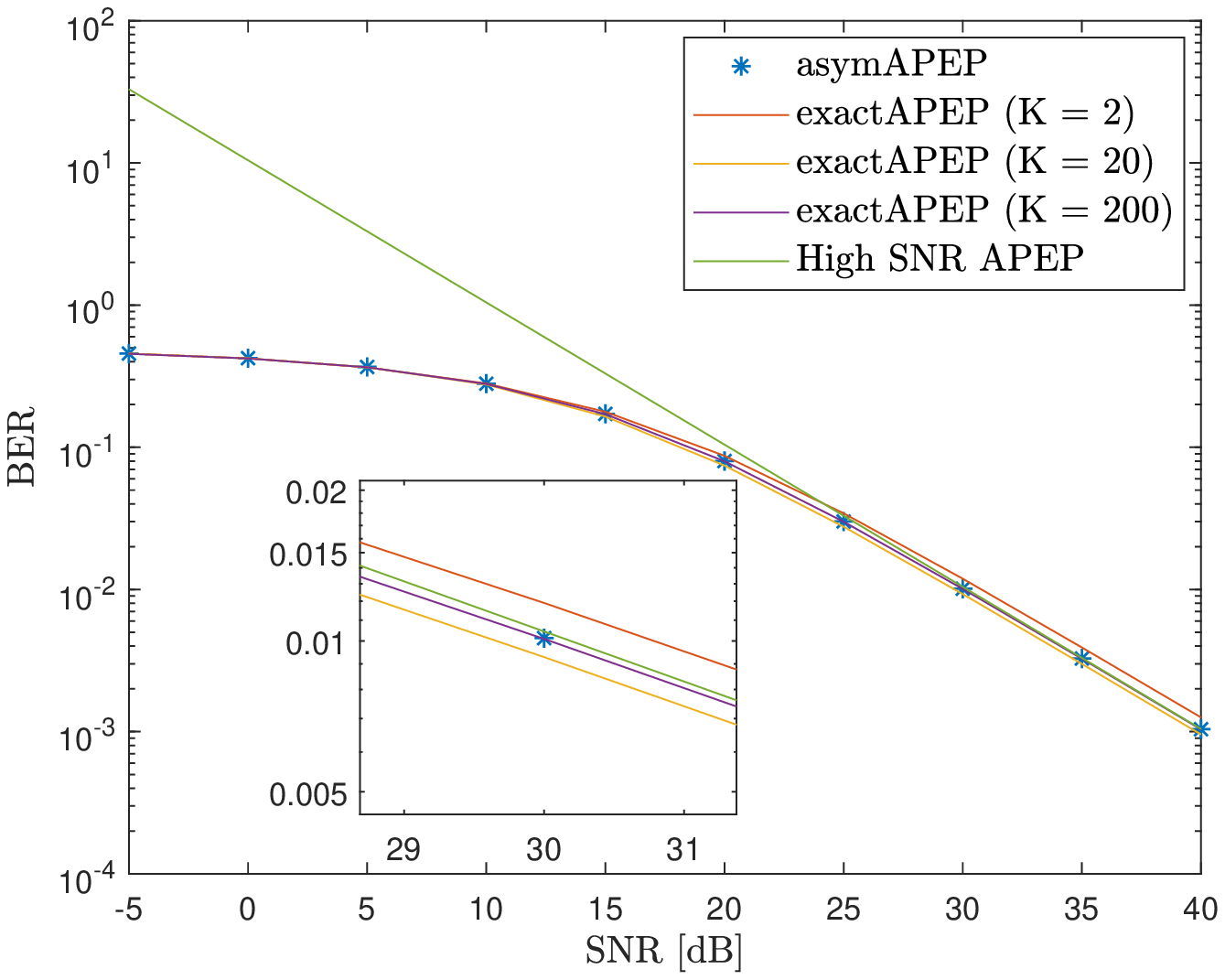}
\caption{ABEP of RecAnt-SM ($P = 2$, $N_r = 1$, $M = 2$ (BPSK)), as a function of $K$. The results are obtained by using $\text{RP}_1$ and $\text{RP}_2$.}
\label{fig:effectK}
\end{figure}
In Fig. \ref{fig:exactABEP}, we compare Monte Carlo simulations against the proposed analytical framework of the bit error probability. In Fig. \ref{fig:effectK}, we study the impact of $K$ on the error probability with the aim of assessing the accuracy of the asymptotic framework for $K \to \infty$. Both figures confirm that our analytical frameworks are accurate, and in agreement with Monte Carlo simulations.

\begin{figure}[!t]
\centering
    \includegraphics[width=0.8\linewidth]{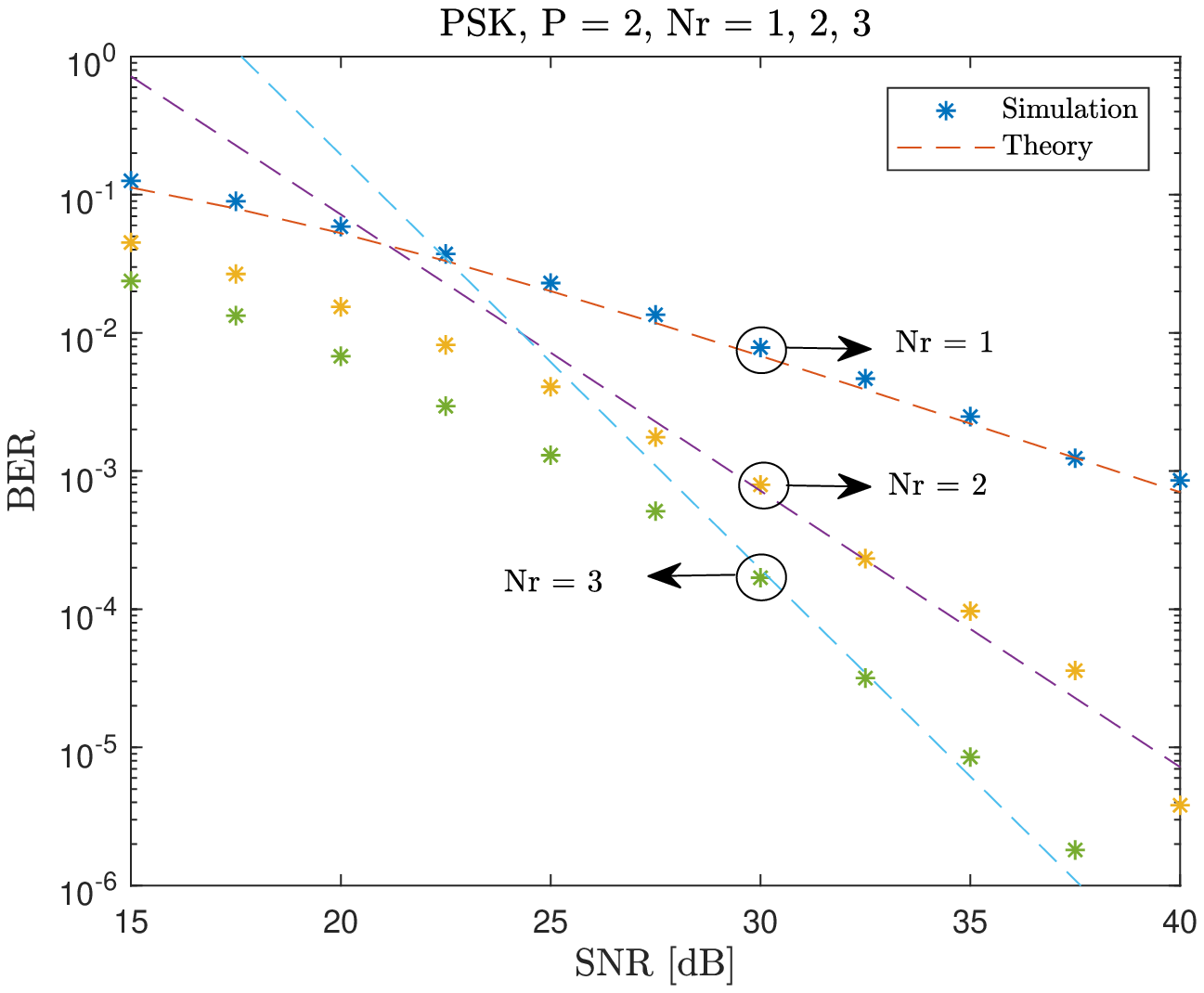}
\caption{ABEP of RecAnt-SM ($P = 2$, $N_r = 1, 2, 3$, $M = 2$ (BPSK)). The results are obtained by using $\text{RP}_1$ and $\text{RP}_2$.}
\label{fig:PSKP2Nr123Prac}
\end{figure}
\begin{figure}[!t]
\centering
    \includegraphics[width=0.8\linewidth]{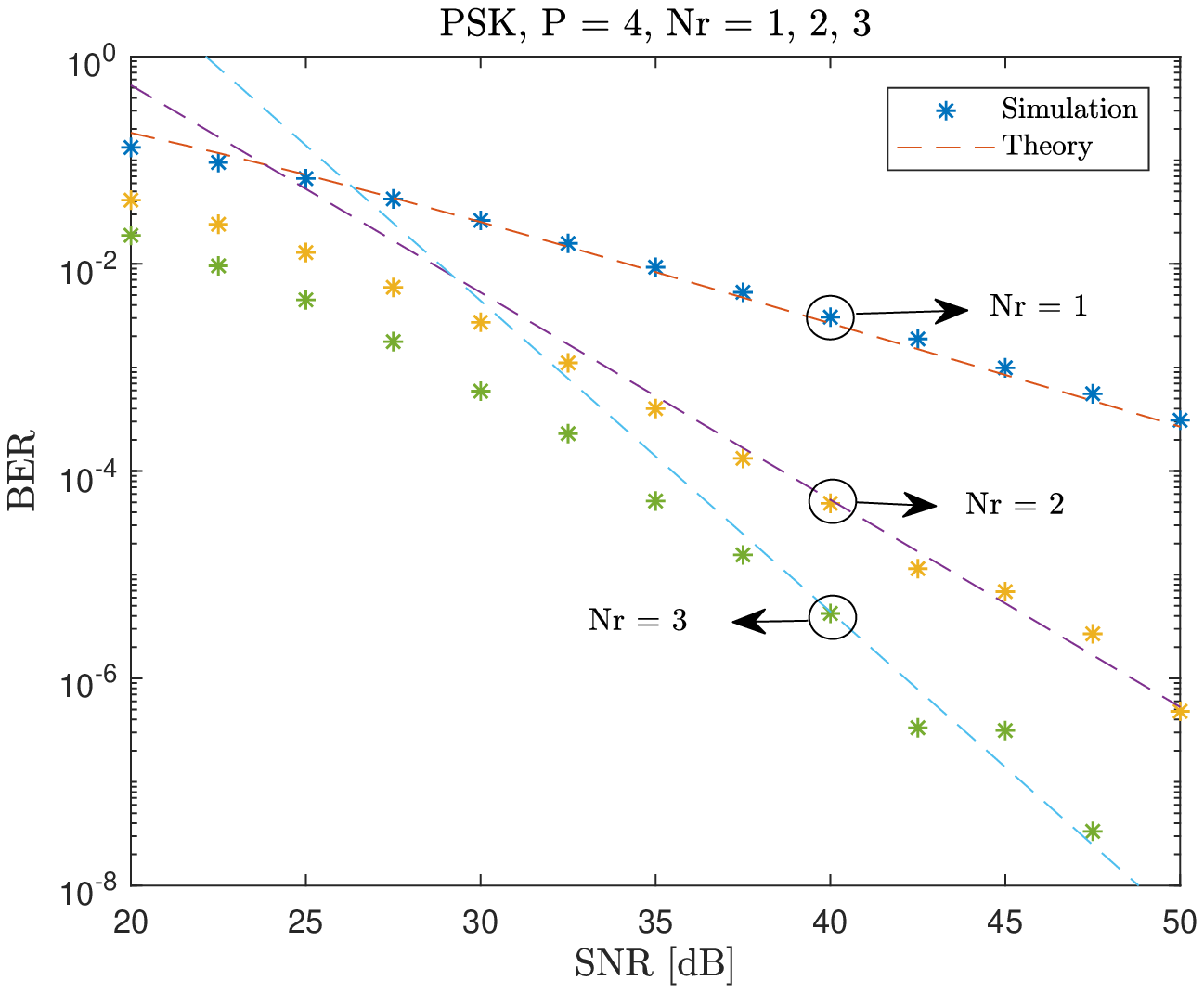}
\caption{ABEP of RecAnt-SM ($P = 4$, $N_r = 1, 2, 3$, $M = 2$ (BPSK)). The results are obtained by using $\text{RP}_1$-$\text{RP}_4$.}
\label{fig:PSKP4Nr123Prac}
\end{figure}
\begin{figure}[!t]
\centering
    \includegraphics[width=0.8\linewidth]{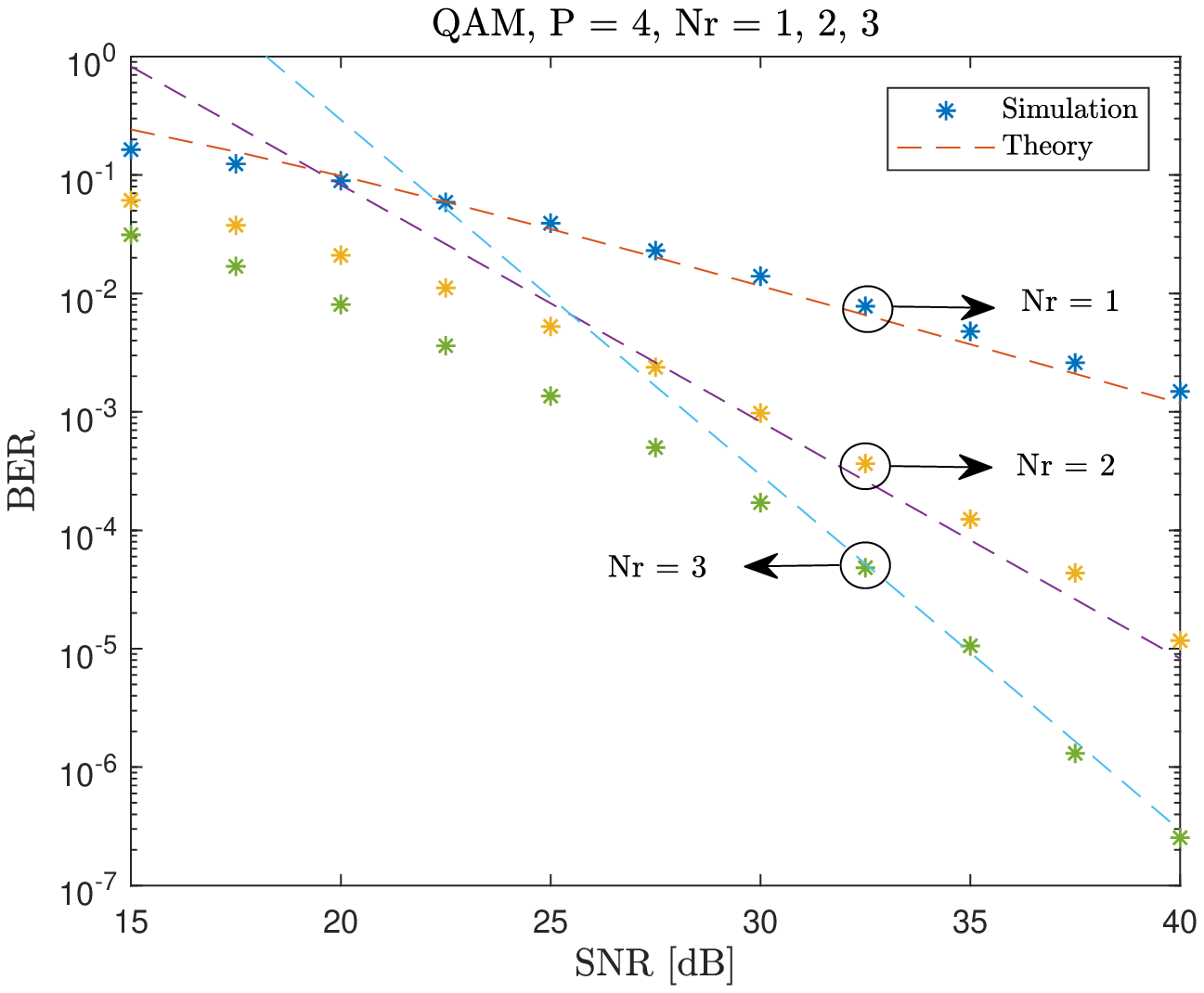}
\caption{ABEP of RecAnt-SM ($P = 4$, $N_r = 1, 2, 3$, $M = 4$ (QPSK)). The results are obtained by using $\text{RP}_1$-$\text{RP}_4$.}
\label{fig:QAMP4Nr123Prac}
\end{figure}
In Figs. \ref{fig:PSKP2Nr123Prac}, \ref{fig:PSKP4Nr123Prac}, and \ref{fig:QAMP4Nr123Prac}, we report the bit error probability as a function of the number of antennas at the receiver. We observe the good accuracy of the proposed analytical frameworks, and, we note, in particular, that RectAnt-SM provides one with a diversity order equal to the number of antennas at the receiver.

\begin{figure}[!t]
\centering
\includegraphics[scale=0.65]{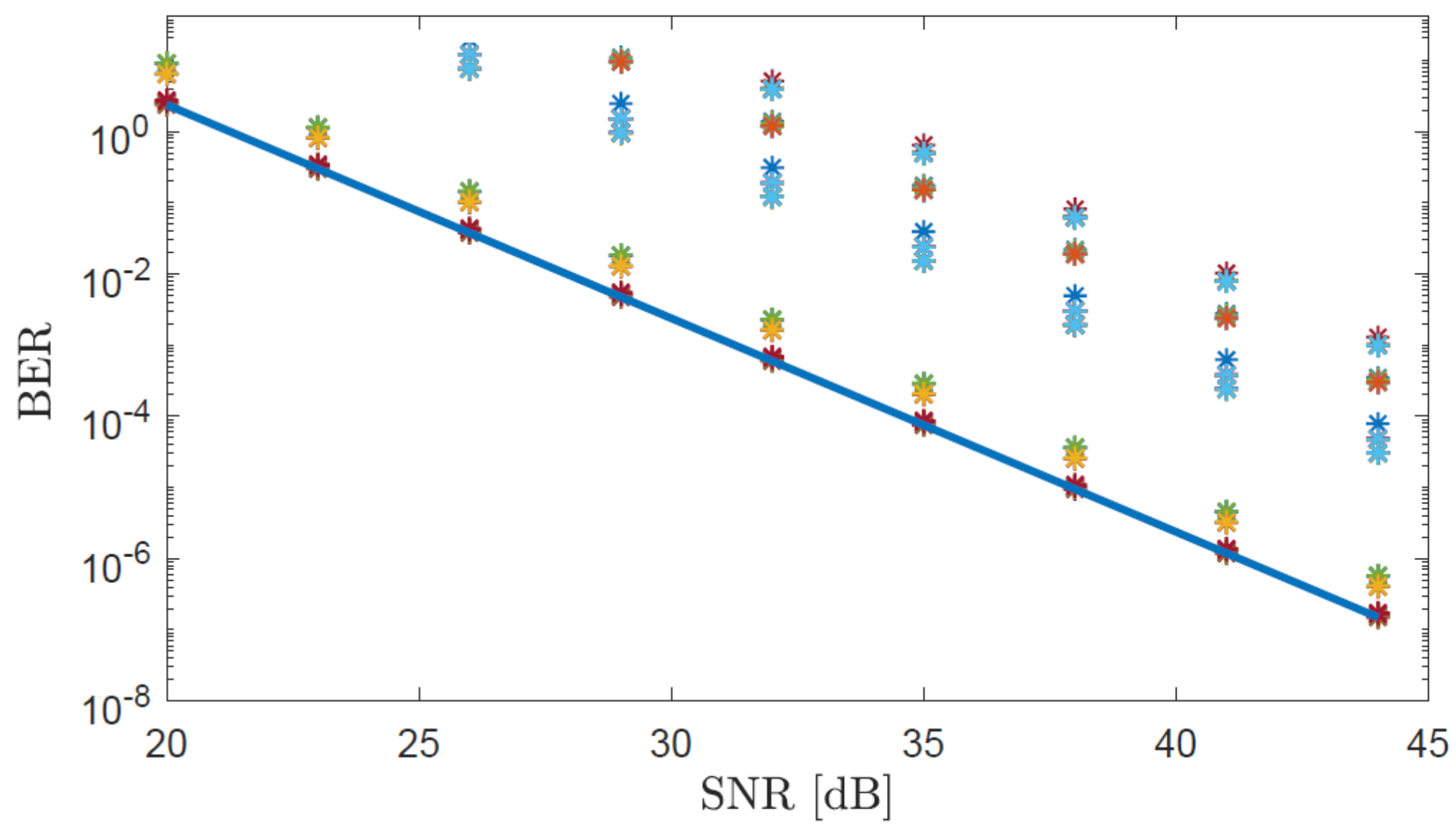}
\caption{ABEP of RecAnt-SSK ($P = 4$, $N_r = 3$). The different markers show the ABEP by choosing 4 RPs out of 8 RPs in order to identify those the offer the best performance.}
\label{fig:illustBestHSNRPRA}
\end{figure}
In Fig. \ref{fig:illustBestHSNRPRA}, we report the APEP as a function of the RPs. In particular, with the aid of the analytical framework of the error probability, we compute the ABEP for all possible combinations of four out of eight RPs in order to identify the impact of the RPs on the error performance. We observe that an appropriate choice of the RPs can yield a significant performance gain. The reason is that, due to practical design constraints, it may not be possible to design several RPs that are very different from each other. Therefore, identifying the best of them that provide good performance as a function of the channel model is an important optimization problem. The proposed analytical framework allows us to solve this optimization problem at a low complexity and high efficiency.

\section{Conclusion}
In this paper, we have studied the performance of spatial modulation based on reconfigurable antennas. We have introduced an analytical framework to compute the error probability, and have described the prototype of a reconfigurable antenna that is specifically designed for application to spatial modulation. By using the radiation patterns obtained from the manufactured antenna prototype, we have shown that spatial modulation based on reconfigurable antennas works in practice, and that its performance can be optimized by appropriately choosing the radiation patterns that minimize the proposed analytical framework of the error probability.


\section{Appendix}

\subsection{Proof of Proposition \ref{pro:APEP2x1} and Proposition \ref{pro:APEP2x12}}
\label{subsec:Pro1}

\subsubsection{Proof of Proposition \ref{pro:APEP2x1}}
Let ${\nu _{{n_r}}} = {\H_{{n_r},q}}{x_n} - {\H_{{n_r},p}}{x_m}$ for ${n_r} = 1, \ldots ,{N_r}$. Recall that $N_r = 1$. We note that ${\nu _{{1}}}$ is a zero mean complex Gaussian variable with variance:
\begin{align}
{\sigma _{{\nu _1}}^2 = \frac{1}{K}\sum\limits_{k = 1}^K {{{\left| {\sqrt {{G_q}(\theta _k^t,\phi _k^t)} \exp (j{\Omega _q}(\theta _k^t,\phi _k^t)){x_n} - \sqrt {{G_p}(\theta _k^t,\phi _k^t)} \exp (j{\Omega _p}(\theta _k^t,\phi _k^t)){x_m}} \right|}^2}} }
\end{align}

Thus, ${\gamma _{p,q,{x_m},{x_n}}} = \mid {\nu _1}{\mid ^2}$ is an exponential random variable whose probability density function is:
\begin{align*}
f_x(x;\lambda) = \lambda\exp(-\lambda x), ~x\geqslant 0
\end{align*}
with $\lambda  = 1/\sigma _{{\nu _1}}^2$.

The APEP can then be formulated as follows:
\begin{align} \label{eq:APEPsigv1}
\APEP = \frac{1}{\pi }\int_0^{\pi /2} {{\Exx_{\sigma _{{\nu _1}}^2}}} \left\{ {{{\left( {1 + \frac{{\rho \sigma _{{\nu _1}}^2}}{{4{{\sin }^2}(\vartheta )}}} \right)}^{ - 1}}} \right\}{\mkern 1mu} d\vartheta
\end{align}

We note that $\sigma _{{\nu _1}}^2$ is a random variable that depend on $\left\{ {\theta _k^t,\phi _k^t} \right\}_{k = 1}^K$. The expectation can be computed by using the approach introduced in \cite{MMGSC2013}, as follows:
\begin{align} \label{eq:APEP2x1Temp}
\APEP = \frac{1}{\pi }\int_0^{\pi /2} {\left( {\int_0^\infty  {{M_S}} (z){\mkern 1mu} dz} \right)} {\mkern 1mu} d\vartheta
\end{align}
where $S = 1 + \frac{{\rho \sigma _{{\nu _1}}^2}}{{4{{\sin }^2}(\vartheta )}}$, and:
\begin{align}
{M_S}(z) = \exp ( - z){M_{\sigma _{{\nu _1}}^2}}\left( {z\frac{\rho }{{4{{\sin }^2}(\vartheta )}}} \right)
\end{align}

With the aid of some algebraic manipulations, we obtain the following:
\begin{align}
\label{eq:MGFsig2v1}
&{M_{\sigma _{{\nu _1}}^2}}\left( {z\frac{\rho }{{4{{\sin }^2}(\vartheta )}}} \right)\notag\\
&= {\Exx_{\theta _k^t,\phi _k^t}}\left\{ {\exp \left( { - z\frac{\rho }{{4K{{\sin }^2}(\vartheta )}}\sum\limits_{k = 1}^K {\psi \left( {p,q,{x_m},{x_n},\theta _k^t,\phi _k^t} \right)} } \right)} \right\} \notag \\
&={\Exx_{\theta _k^t,\phi _k^t}}\left\{ {\prod\limits_{k = 1}^K {\exp \left( { - z\frac{\rho }{{4K{{\sin }^2}(\vartheta )}}\psi \left( {p,q,{x_m},{x_n},\theta _k^t,\phi _k^t} \right)} \right)} } \right\} \notag \\
&=\resizebox{\textwidth}{!}{$\prod\limits_{k = 1}^K {\left( {\int_{ - \pi }^\pi  {\int_0^\pi  {\exp \left( { - z\frac{\rho }{{4K{{\sin }^2}(\vartheta )}}\psi \left( {p,q,{x_m},{x_n},\theta _k^t,\phi _k^t} \right)} \right)} } {f_\theta }(\theta _k^t){f_\phi }(\phi _k^t)d\theta _k^td\phi _k^t} \right)}$} \\
\text{where:} \notag \\
&\psi \left( {p,q,{x_m},{x_n},\theta _k^t,\phi _k^t} \right) = {\left| {\sqrt {{G_q}(\theta _k^t,\phi _k^t)} \exp (j{\Omega _q}(\theta _k^t,\phi _k^t)){x_n} - \sqrt {{G_p}(\theta _k^t,\phi _k^t)} \exp (j{\Omega _p}(\theta _k^t,\phi _k^t)){x_m}} \right|^2} \notag
\end{align}

The proof follows from the following identity:
\begin{align}
{\int_{ - \pi }^\pi  {\int_0^\pi  {{f_\theta }({\theta ^t}){f_\phi }({\phi ^t})d\theta _k^td\phi _k^t = } } \int_{ - \pi }^\pi  {{f_\theta }({\theta ^t})d\theta _k^t\int_0^\pi  {{f_\phi }({\phi ^t})d\phi _k^t = } } 1}
\end{align}

\subsubsection{Proof of Proposition \ref{pro:APEP2x12}}
From (\ref{eq:APEPsigv1}), we obtain, in the high-SNR regime, the following:
\begin{align}
\APEP_{p,q} &= \Exx_{\sigma^2_{\nu_1}}\left\{\frac{1}{\pi} \int_0^{\pi/2}\left(1+\frac{\rho\sigma^2_{\nu_1}}{4\sin^2(\vartheta)}\right)^{-1}\,d\vartheta \right\} \nonumber\\
			&\overset{\rho \gg 1}{\leq} \Exx_{\sigma^2_{\nu_1}}\left\{ \frac{1}{\pi} \int_0^{\pi/2}\left(\frac{\rho\sigma^2_{\nu_1}}{4\sin^2(\vartheta)}\right)^{-1}\,d\vartheta \right\} \nonumber\\
			&= \frac{1}{\rho}\Exx_{\sigma^2_{\nu_1}}\left\{\frac{1}{\sigma^2_{\nu_1}} \left( \underbrace{\frac{1}{\pi} \int_0^{\pi/2}\left(\frac{1}{4\sin^2(\vartheta)}\right)^{-1}\,d\vartheta}_{=1} \right) \right\} \nonumber\\
			&= \frac{1}{\rho}\Exx_{\sigma^2_{\nu_1}}\left\{\frac{1}{\sigma^2_{\nu_1}}\right\} \label{eq:APEPSNRBound1}
\end{align}
\label{subsec:Pro2}

The rest of the proof follows by using similar steps as for the proof of Proposition $\ref{pro:APEP2x1}$.

\subsection{Proof of Proposition \ref{pro:limK} and Proposition \ref{pro:limK1}}

\subsubsection{Proof of Proposition \ref{pro:limK}}
By using the Maclaurin series expansion and keeping the first two dominant terms, we have the following:
\begin{align}
\exp \left( { - z\bar \zeta \left( {K,\vartheta } \right)\psi \left( {p,q,{x_m},{x_n},{\theta ^t},{\phi ^t}} \right)} \right) &= 1 - z\bar \zeta \left( {K,\vartheta } \right)\psi \left( {p,q,{x_m},{x_n},{\theta ^t},{\phi ^t}} \right) + \notag\\
&\qquad\:\:\:\:{\cal O}\left( {z\bar \zeta \left( {K,\vartheta } \right)\psi \left( {p,q,{x_m},{x_n},{\theta ^t},{\phi ^t}} \right)} \right)\notag\\
&\approx 1 - z\bar \zeta \left( {K,\vartheta } \right)\psi \left( {p,q,{x_m},{x_n},{\theta ^t},{\phi ^t}} \right)
\end{align}

By using this approximation, we have the following:
\begin{align}
&\int_0^\pi  {\int_{ - \pi }^\pi  {\exp } } \left( { - z\bar \zeta \left( {K,\vartheta } \right)\psi \left( {p,q,{x_m},{x_n},{\theta ^t},{\phi ^t}} \right)} \right){f_\theta }({\theta ^t}){f_\phi }({\phi ^t}){\mkern 1mu} d{\theta ^t}{\mkern 1mu} d{\phi ^t}\notag \\
&\approx \int_0^\pi  {\int_{ - \pi }^\pi  {\left( {1 - z\bar \zeta \left( {K,\vartheta } \right)\psi \left( {p,q,{x_m},{x_n},{\theta ^t},{\phi ^t}} \right)} \right)} } {f_\theta }({\theta ^t}){f_\phi }({\phi ^t}){\mkern 1mu} d{\theta ^t}{\mkern 1mu} d{\phi ^t}\notag \\
& = 1 - z\bar \zeta \left( {K,\vartheta } \right)\int_0^\pi  {\int_{ - \pi }^\pi  {\psi \left( {p,q,{x_m},{x_n},{\theta ^t},{\phi ^t}} \right)} } {f_\theta }({\theta ^t}){f_\phi }({\phi ^t}){\mkern 1mu} d{\theta ^t}{\mkern 1mu} d{\phi ^t}
\end{align}

Let us consider the following integral:
\begin{align}
\Theta  = \int_0^\pi  {\int_{ - \pi }^\pi  {\psi \left( {p,q,{x_m},{x_n},{\theta ^t},{\phi ^t}} \right)} } {f_\theta }({\theta ^t}){f_\phi }({\phi ^t})d{\theta ^t}d{\phi ^t}
\end{align}

In addition, the following holds true:
\begin{align}
\mathop {\lim }\limits_{K \to  + \infty } {\rm{ }}{\left( {1 - \frac{{z\frac{\rho }{{4{{\sin }^2}(\vartheta )}}\Theta }}{K}} \right)^K} = {\rm{ }}\exp \left( { - z\frac{\rho }{{4{{\sin }^2}(\vartheta )}}\Theta } \right)
\end{align}
where we used the following notable limit:
\begin{align} \label{eq:spLim}
\mathop {\lim }\limits_{x \to  + \infty } {\left( {1 + \frac{k}{x}} \right)^x} = {e^k},
\end{align}

Therefore, the asymptotic APEP is:
\begin{align} \label{eq:asymAPEPapp}
\APEP &\leq \frac{1}{\pi }\int_0^{\pi /2} {\int_0^\infty  {\exp \left( { - z\left( {1 + \frac{\rho }{{4{{\sin }^2}(\vartheta )}}\Theta } \right)} \right)} } dzd\vartheta \notag\\
&{\mathop  = \limits^{(a)} \frac{1}{\pi }\int_0^{\pi /2} {{{\left( {1 + \frac{\rho }{{{4{\sin }^2}(\vartheta )}}\Theta } \right)}^{ - 1}}} d\vartheta } \notag\\
&{\mathop  = \limits^{(b)} \frac{1}{2}\left( {1 - \sqrt {\frac{{\rho \Theta }}{{\rho \Theta  + 1}}} } \right)}
\end{align}
where the first equality (a) comes the fact that
$\int_0^\infty  {\exp \left( { - zc} \right) = } \frac{1}{c},c > 0$
and the second equality (b) follows from (5A.4a) in \cite{simon2005digital}. This concludes the proof.

\subsubsection{Proof of Proposition \ref{pro:limK1}}

By using steps similar to Proposition \ref{pro:limK}, the asymptotic APEP can be written as follows:
\begin{align}
\APEP &\leq \frac{1}{\pi }\int_0^{\pi /2} {{{\left( {1 + \frac{\rho }{{{4{\sin }^2}(\vartheta )}}\Theta } \right)}^{ - 1}}} d\vartheta   \notag\\
&\mathop  \le \limits^{\rho  \gg 1} \frac{1}{{\rho \Theta }}\left( {\underbrace {\frac{1}{\pi }\int_0^{\pi /2} {{{\left( {\frac{1}{{4{{\sin }^2}(\vartheta )}}} \right)}^{ - 1}}} {\mkern 1mu} d\vartheta }_{ = 1}} \right),
\end{align}
which concludes the proof.

\subsection{Proof of Proposition \ref{pro:APEP2x2}}
\label{subsec:Pro3}
We first introduce the following lemma \cite{Turin1960} for application to hermitian quadratic forms in complex normal variables.
\begin{lemma} \label{lm:MGFQ}
Let $v_n (n = 1, \ldots, N)$ be a set of complex Gaussian random variables having zero mean. Let $\kappa$, with $\v = [v_1, \cdots, v_N]^T$, be an Hermitian quadratic form:
\begin{align}
\kappa = \v^H\I_N\v
\end{align}

Its MGF is as follows:
\begin{align}
M_{\kappa}(s) = \prod_{n = 1}^N (1-s\lambda_n)^{-1}
\end{align}
where $\lambda_n$ is the $n$th eigenvalue of the covariance matrix $\R_{\v} = \Exx\{\v\v^H\}$.
\end{lemma}
\begin{proof}
See \cite{Turin1960}.
\end{proof}

The proof of Proposition \ref{pro:APEP2x2} can be split in three steps.\\

\textit{Step 1}: By using Lemma \ref{lm:MGFQ} with $N = N_r = 2$, we have:
\begin{align} \label{eq:APEPrho2}
\APEP \approx \frac{1}{{{\rho ^2}}}{\Exx_{\theta _k^t,\phi _k^t,\theta _k^r,\phi _k^r}}\left\{ {\left( {\frac{3}{{{\lambda _1}{\lambda _2}}}} \right)} \right\}
\end{align}
where $\lambda_1$ and $\lambda_2$ are eigenvectors of the covariance matrix:
\begin{align} \label{eq:R}
\R = \left[ {\begin{array}{*{20}{c}}
{\Exx\left\{ {\left. {{\nu _1}{\nu ^*}_1} \right\}} \right.}&{\Exx\left\{ {\left. {{\nu _1}{\nu ^*}_2} \right\}} \right.}\\
{\Exx\left\{ {\left. {{\nu _2}{\nu ^*}_1} \right\}} \right.}&{\Exx\left\{ {\left. {{\nu _2}{\nu ^*}_2} \right\}} \right.}
\end{array}} \right]
\end{align}
and $\R$ is assumed to be full rank.

It is worth mentioning that $\lambda_1$ and $\lambda_2$ depend on the random variables ${\theta _k^t,\phi _k^t,\theta _k^r,\phi _k^r}$. In particular, from Lemma~\ref{lm:MGFQ}, we have:
\begin{align} \label{eq:APEP2x2E}
\APEP = \Exx_{\theta _k^t,\phi _k^t,\theta _k^r,\phi _k^r}\left\{ {\left. {\frac{1}{\pi }\int\limits_0^{\pi /2} {\prod\limits_{{n_r} = 1}^2 {{{(1 + \frac{\rho }{{4{{\sin }^2}\left( \vartheta  \right)}}{\lambda _{n_r}})}^{ - 1}}} d\vartheta } } \right\}} \right.
\end{align}

Also, we have the following:
\begin{align}
\frac{1}{{\left( {1 + \frac{\rho }{{4{{\sin }^2}\left( \vartheta  \right)}}{\lambda _1}} \right)\left( {1 + \frac{\rho }{{4{{\sin }^2}\left( \vartheta  \right)}}{\lambda _2}} \right)}} =
\frac{{{A_1}}}{{\left( {1 + \frac{\rho }{{4{{\sin }^2}\left( \vartheta  \right)}}{\lambda _1}} \right)}} + \frac{{{A_2}}}{{\left( {1 + \frac{\rho }{{4{{\sin }^2}\left( \vartheta  \right)}}{\lambda _2}} \right)}}
\end{align}
which yields:
\begin{align}
{A_1} = \frac{{{\lambda _1}}}{{{\lambda _1} - {\lambda _2}}}, \qquad
{A_2} =  - \frac{{{\lambda _2}}}{{{\lambda _1} - {\lambda _2}}} \label{eq:A12}
\end{align}

For high-SNR, we have:
\begin{align}
{\left( {1 + \frac{\rho }{{4{{\sin }^2}\left( \vartheta  \right)}}{\lambda _{n_r}}} \right)^{ - 1}} \approx
{\left( {\frac{\rho }{{4{{\sin }^2}\left( \vartheta  \right)}}{\lambda _{n_r}}} \right)^{ - 1}} - {\left( {\frac{\rho }{{4{{\sin }^2}\left( \vartheta  \right)}}{\lambda _{n_r}}} \right)^{ - 2}} \label{eq:2Appr}
\end{align}
where we have used the second-order Taylor approximation.

Thus, by substituting (\ref{eq:A12}) and (\ref{eq:2Appr}) in (\ref{eq:APEP2x2E}), we obtain:
\begin{align}
\APEP &= {\Exx_{\theta _k^t,\phi _k^t,\theta _k^r,\phi _k^r}}\left\{ {\left. {\sum\limits_{i = 1}^2 {{A_i}\left[ {\frac{1}{\pi }\int\limits_0^{\pi /2} {{{\left( {\frac{\rho }{{4{{\sin }^2}\left( \vartheta  \right)}}{\lambda _i}} \right)}^{ - 1}}d\vartheta  - \frac{1}{\pi }\int\limits_0^{\pi /2} {{{\left( {\frac{\rho }{{4{{\sin }^2}\left( \vartheta  \right)}}{\lambda _i}} \right)}^{ - 2}}d\vartheta } } } \right]} } \right\}} \right. \notag\\
&= {\Exx_{\theta _k^t,\phi _k^t,\theta _k^r,\phi _k^r}}\left\{ {\left. {\frac{1}{\rho }\left[ {\frac{1}{{{\lambda _i}}}\frac{1}{\pi }\int\limits_0^{\pi /2} {{{\left( {\frac{1}{{4{{\sin }^2}\vartheta }}} \right)}^{ - 1}}d\omega }  - \frac{1}{\rho }\frac{1}{{{\lambda ^2}_i}}\frac{1}{\pi }\int\limits_0^{\pi /2} {{{\left( {\frac{1}{{4{{\sin }^2}\left( \vartheta  \right)}}} \right)}^{ - 2}}d\vartheta } } \right]} \right\}} \right. \notag\\
&\mathop  = \limits^{(a)} { \Exx_{\theta _k^t,\phi _k^t,\theta _k^r,\phi _k^r}}\left\{ {\left. {\frac{1}{\rho }\left( {\frac{{{\lambda _1}}}{{{\lambda _1} - {\lambda _2}}}\left[ {\frac{1}{{{\lambda _1}}} - \frac{3}{{\rho {\lambda ^2}_1}}} \right] - \frac{{{\lambda _2}}}{{{\lambda _1} - {\lambda _2}}}\left[ {\frac{1}{{{\lambda _2}}} - \frac{3}{{\rho {\lambda ^2}_2}}} \right]} \right)} \right\}} \right. \notag\\
&= {\Exx_{\theta _k^t,\phi _k^t,\theta _k^r,\phi _k^r}}\left\{ {\frac{1}{{{\rho ^2}}}\left( {\frac{3}{{{\lambda _1}{\lambda _2}}}} \right)} \right\}
\end{align}
where (a) follows from:
\begin{align}
\frac{1}{\pi }\int\limits_0^{\pi /2} {{{\left( {\frac{1}{{4{{\sin }^2}\left( \omega  \right)}}} \right)}^{ - 1}}d\omega }  = 1, \qquad
\frac{1}{\pi }\int\limits_0^{\pi /2} {{{\left( {\frac{1}{{4{{\sin }^2}\left( \omega  \right)}}} \right)}^{ - 2}}d\omega }  = 3
\end{align}

\textit{Step 2}: We compute the explicit expression of the product $\lambda _1 \lambda _2$. To this end, we introduce the following lemma.
\begin{lemma}
Assume that $\R$, defined in (\ref{eq:R}), is full rank and has two distinct eigenvalues ${\lambda _1}$ and ${\lambda _2}$. The product of the two eigenvectors ${\lambda _1}$ and ${\lambda _2}$ is as follows:
\begin{align} \label{eq:lam12}
&{\lambda_1 \lambda _2} = {\Exx_{{\theta ^t},{\phi ^t},\theta _k^r,\phi _k^r}}\left( {\frac{1}{{{K^2}}}\left[ {\sum\limits_{k = 1}^K {\sum\limits_{k' = 1}^K {\chi \left( {{\theta _k}^t,\phi _k^t} \right)\chi \left( {{\theta _{k'}}^t,\phi _{k'}^t} \right)} } \Upsilon} \right]} \right)
\end{align}
where
\begin{align}
\chi \left( {{\theta _k}^t,\phi _k^t} \right) = &\sqrt {{G_q}\left( {{\theta _k}^t,\phi _k^t} \right)} \exp \left( {j{\Omega _q}\left( {{\theta _k}^t,\phi _k^t} \right)} \right) - \sqrt {{G_p}\left( {{\theta _k}^t,\phi _k^t} \right)} \exp \left( {j{\Omega _p}\left( {{\theta _k}^t,\phi _k^t} \right)} \right)
\end{align}
and
\begin{align}
\Upsilon ={1 - \cos \left( {\left\| \k \right\|{d}\left( {\sin \left( {{\theta ^r}} \right)\sin \left( {{\phi ^r}} \right) - \sin \left( {\theta {'^r}} \right)\sin \left( {\phi {'^r}} \right)} \right)} \right)}
\end{align}
\end{lemma}
\begin{proof}
By definition:
\begin{align}
\nu _1 &= \frac{1}{{\sqrt K }}\sum\limits_{k = 1}^K {{\beta _k}\chi \left( {{\theta _k}^t,\phi _k^t} \right)} \label{eq:nu1}\\
{\nu _2} &= \frac{1}{{\sqrt K }}\sum\limits_{k = 1}^K {\beta _k}{\chi \left( {{\theta _k}^t,\phi _k^t} \right)\exp \left( {j\left\| k \right\|{d}\sin \left( {\theta _k^r} \right)\sin \left( {\phi _k^r} \right)} \right)} \label{eq:nu2}
\end{align}
Moreover, it is known that the product of the eigenvalues is equal to the determinant of the covariance matrix (i.e., $\det (\R) = {\lambda _1}{\lambda _2}$). Thus, ${\lambda _1}{\lambda _2}$ can be computed directly from $\det (\R)$ as follows:
\begin{align} \label{eq:detR2}
\det \left( \R \right) = \Exx\left\{ {\left. {{\nu _1}{\nu ^*}_1} \right\}} \right.\Exx\left\{ {\left. {{\nu _2}{\nu ^*}_2} \right\}} \right. -
\Exx\left\{ {\left. {{\nu _2}{\nu ^*}_1} \right\}} \right.\Exx\left\{ {\left. {{\nu _1}{\nu ^*}_2} \right\}} \right.
\end{align}
By substituting (\ref{eq:nu1}) and (\ref{eq:nu2}) into (\ref{eq:detR2}), we obtain (\ref{eq:lam12}) where we have used the identity: \[\frac{{\exp \left( {j\phi } \right) + \exp \left( { - j\phi } \right)}}{2} = \cos \left( \phi  \right)\] and if $k = k'$ we have $ {1 - \exp \left( {j\left\| {{\k}} \right\|{d}\left( {\sin \left( {\theta _k^r} \right)\sin \left( {\phi _k^r} \right) - \sin \left( {\theta _{k'}^r} \right)\sin \left( {\phi _{k'}^r} \right)} \right)} \right)}  = 0$.
\end{proof}

\textit{Step 3:} We exploit the asymptotic analysis introduced in Proposition \ref{pro:limK} to derive a closed-form expression of the APEP. We apply the following formula that is special case of (40) in \cite{MMGSC2013} for $X = {{{\lambda _1}{\lambda _2}}}$:
\begin{align} \label{eq:Exm1}
\Exx\left( {{X^{ - 1}}} \right) = \int_0^\infty  {{M_X}\left( { - z} \right)dz}
\end{align}
where ${M_X}\left( z \right) = \Exx_X\left\{\exp \left( {Xz} \right) \right\}$.

Then, we obtain:
\begin{align} \label{eq:ExXz}
&{\Exx_X}\left( {\exp \left( {Xz} \right)} \right){\rm{ }} = {\Exx_{{\theta ^t},{\phi ^t},{\theta ^r},{\phi ^r}}}\left( {\prod\limits_{k = 1}^K {\prod\limits_{k' = 1}^K {\exp } } \left( {\frac{1}{{{K^2}}}{\chi}\left( {{\theta _k}^t,\phi _k^t} \right){\chi}\left( {{\theta _{k'}}^t,\phi _{k'}^t} \right) \Upsilon z} \right)} \right)\notag\\
&=\resizebox{\textwidth}{!}{$\prod\limits_{k = 1}^K {\prod\limits_{k' = 1}^K {\left( {\int_{ - \pi }^\pi  {\int_{ - \pi }^\pi  {\int_{ - \pi }^\pi  {\int_{ - \pi }^\pi  {\int_0^\pi  {\int_0^\pi  {\int_0^\pi  {\int_0^\pi  {\exp \left( {\frac{1}{{{K^2}}}{\chi}\left( {{\theta _k}^t,\phi _k^t} \right){\chi}\left( {{\theta _{k'}}^t,\phi _{k'}^t} \right) \Upsilon z} \right){f_{{\phi ^t}_{k'}}}{f_{{\phi ^t}_k}}{f_{{\phi ^r}_{k'}}}{f_{{\phi ^r}_k}}{f_{{\theta ^t}_{k'}}}{f_{{\theta ^t}_k}}{f_{{\theta ^r}_{k'}}}{f_{{\theta ^r}_k}}d{\phi _{k'}}^td{\phi _k}^td{\phi _{r'}}^rd{\phi _k}^r} } } } } } } } d{\theta _{k'}}^td{\theta _k}^td{\theta _{r'}}^rd{\theta _k}^r} \right)} }$}  \notag\\
&=\resizebox{\textwidth}{!}{${\left( {\int_{ - \pi }^\pi  {\int_{ - \pi }^\pi  {\int_{ - \pi }^\pi  {\int_{ - \pi }^\pi  {\int_0^\pi  {\int_0^\pi  {\int_0^\pi  {\int_0^\pi  {\exp \left( {\frac{1}{{{K^2}}}{\chi}\left( {{\theta ^t},{\phi ^t}} \right)\chi\left( {\theta {'^t},\phi {'^t}} \right)\Upsilon z} \right){f_{{\phi ^t}}}{f_{\phi {'^t}}}{f_{{\phi ^r}}}{f_{\phi {'^r}}}{f_{{\theta ^t}}}{f_{\theta {'^t}}}{f_{{\theta ^r}}}{f_{\theta {'^r}}}d{\phi ^t}d\phi {'^t}d{\phi ^r}d\phi {'^r}} } } } } } } } d{\theta ^t}d\theta {'^t}d{\theta ^r}d\theta {'^r}} \right)^{{K^2}}}$}
\end{align}
where the last equality follows from the assumption of independent and identically distributed random variables.

Moreover, by using the following approximation:
\begin{align}
\exp \left( {\frac{1}{{{K^2}}}{\chi}\left( {{\theta ^t},{\phi ^t}} \right)\chi\left( {\theta {'^t},\phi {'^t}} \right)\Upsilon z} \right) \approx
1 + \frac{1}{{{K^2}}}{\chi}\left( {{\theta ^t},{\phi ^t}} \right)\chi\left( {\theta {'^t},\phi {'^t}} \right)\Upsilon z
\end{align}
in (\ref{eq:ExXz}) and using the notable limit in (\ref{eq:spLim}), we obtain the following:
\begin{align} \label{eq:EXKinf}
&{\Exx_X}\left( {\exp \left( {Xz} \right)} \right) \mathop  = \limits^{K \to \infty } \notag\\
&\resizebox{\textwidth}{!}{$\exp \left( {z\int_{ - \pi }^\pi  {\int_{ - \pi }^\pi  {\int_{ - \pi }^\pi  {\int_{ - \pi }^\pi  {\int_0^\pi  {\int_0^\pi  {\int_0^\pi  {\int_0^\pi  {\left( {{\chi}\left( {{\theta ^t},{\phi ^t}} \right)\chi\left( {\theta {'^t},\phi {'^t}} \right)\Upsilon} \right){f_{{\phi ^t}}}{f_{\phi {'^t}}}{f_{{\phi ^r}}}{f_{\phi {'^r}}}{f_{{\theta ^t}}}{f_{\theta {'^t}}}{f_{{\theta ^r}}}{f_{\theta {'^r}}}d{\phi ^t}d\phi {'^t}d{\phi ^r}d\phi {'^r}} } } } } } } } d{\theta ^t}d\theta {'^t}d{\theta ^r}d\theta {'^r}} \right)$}
\end{align}

Finally, substituting (\ref{eq:EXKinf}) in (\ref{eq:Exm1}), and then (\ref{eq:Exm1}) in (\ref{eq:APEPrho2}), we conclude the proof by using the following result:
\begin{align}
&\resizebox{\textwidth}{!}{$\int_{ - \pi }^\pi  {\int_0^\pi  {\int_{ - \pi }^\pi  {\int_0^\pi  {\left( {1 - \cos \left( {\left\| \k \right\|{d}\left( {\sin \left( {{\theta ^r}} \right)\sin \left( {{\phi ^r}} \right) - \sin \left( {\theta {'^r}} \right)\sin \left( {\phi {'^r}} \right)} \right)} \right)} \right)} } } } {f_{{\phi ^r}}}{f_{\phi {'^r}}}{f_{{\theta ^r}}}{f_{\theta {'^r}}}d{\phi ^r}d{\theta ^r}d\phi {'^r}d\theta {'^r} =$} \notag\\
&\resizebox{\textwidth}{!}{$1 - {\left( {\int_{ - \pi }^\pi  {\int_0^\pi  {\cos \left( {\left\| \k \right\|{d}\left( {\sin \left( {{\theta ^r}} \right)\sin \left( {{\phi ^r}} \right)} \right)} \right)} {f_{{\phi ^r}}}{f_{{\theta ^r}}}d{\phi ^r}d{\theta ^r}} } \right)^2} - {\left( {\int_{ - \pi }^\pi  {\int_0^\pi  {\sin \left( {\left\| \k \right\|{d}\left( {\sin \left( {{\theta ^r}} \right)\sin \left( {{\phi ^r}} \right)} \right)} \right){f_{{\phi ^r}}}{f_{{\theta ^r}}}d{\phi ^r}d{\theta ^r}} } } \right)^2}$}
\end{align}

\subsection{Proof of Theorem 1}
\label{subsec:Theo1}
The proof of Theorem \ref{the:Nr} generalizes the steps of Proposition \ref{pro:APEP2x2} as follows.

\textit{Step 1}: From Lemma \ref{lm:MGFQ}, we have:
\begin{align}
\APEP = {\Exx_{\theta _k^t,\phi _k^t,\theta _k^r,\phi _k^r}}\left\{ {\left. {\frac{1}{\pi }\int\limits_0^{\pi /2} {\prod\limits_{n = 1}^{{N_r}} {{{(1 + \frac{\rho }{{4{{\sin }^2}\left( \vartheta  \right)}}{\lambda _n})}^{ - 1}}} d\vartheta } } \right\}} \right.
\end{align}

Then, we need to identify the coefficients $\left\{ {{A_{{n_r}}}} \right\}_{{n_r} = 1}^{{N_r}}$ so that the following is satisfied:
\begin{align}
\frac{1}{{\prod\limits_{{n_r} = 1}^{{N_r}} {\left( {1 + \frac{\rho }{{4{{\sin }^2}\left( \omega  \right)}}{\lambda _{n_r}}} \right)} }} = \sum\limits_{{n_r} = 1}^{{N_r}} {\frac{{{A_{n_r}}}}{{\left( {1 + \frac{\rho }{{4{{\sin }^2}\left( \omega  \right)}}{\lambda _{n_r}}} \right)}}} \label{eq:An}
\end{align}

To this end, we can exploit the general formula of partial fraction decomposition \cite[p. 66-67]{jeffrey2007table}:
\begin{align}
{A_{n_r}} = \frac{{{\lambda _{n_r}}^{{N_r} - 1}}}{{\prod\limits_{{n_r} \ne l = 1}^{{N_r}} {\left( {{\lambda _{n_r}} - {\lambda _l}} \right)} }}
\end{align}
for $l, n_r = 1, \ldots, N_r$.

Thus, we have:
\begin{align}
\APEP &\approx {\Exx_{_{{\theta ^t},{\phi ^t},{\theta ^r},{\phi ^r}}}}\left\{ {\left. {\frac{1}{\pi }\int\limits_0^{\pi /2} {\sum\limits_{{n_r} = 1}^{{N_r}} {{A_n}\left[ {\sum\limits_{m = 1}^{{N_r}} {{{\left( { - 1} \right)}^{m + 1}}{{\left( {\frac{\rho }{{4{{\sin }^2}\left( \vartheta  \right)}}{\lambda _{{n_r}}}} \right)}^{ - m}}} } \right]} d\vartheta } } \right\}} \right. \notag\\
 &= {\Exx_{{\theta ^t},{\phi ^t},{\theta ^r},{\phi ^r}}}\left\{ {\sum\limits_{{n_r} = 1}^{{N_r}} {{A_n}\left[ {\sum\limits_{m = 1}^{{N_r}} {{{\left( { - 1} \right)}^{m + 1}}\frac{1}{\pi }\int_0^{\pi /2} {{{\left( {\frac{\rho }{{4{{\sin }^2}\left( \vartheta  \right)}}{\lambda _{{n_r}}}} \right)}^{ - m}}d\vartheta } } } \right]} } \right\} \notag\\
&= {\Exx_{_{{\theta ^t},{\phi ^t},{\theta ^r},{\phi ^r}}}}\left\{ {\sum\limits_{{n_r} = 1}^{{N_r}} {{A_n}\left[ {\sum\limits_{m = 1}^{{N_r}} {{{\left( { - 1} \right)}^{m + 1}}{{\left( {\rho {\lambda _{{n_r}}}} \right)}^{ - m}}\left( {\frac{1}{\pi }\int_0^{\pi /2} {{{\left( {\frac{1}{{4{{\sin }^2}\left( \vartheta  \right)}}} \right)}^{ - m}}d\vartheta } } \right)} } \right]} } \right\}
\end{align}

In the high-SNR regime, we can use the $N_r$th-order Taylor approximation as follows:
\begin{align}\label{eq:NrHighSNR}
{\left( {1 + \frac{\rho }{{4{{\sin }^2}\left( \vartheta  \right)}}{\lambda _{{n_r}}}} \right)^{ - {N_r}}} \approx
\sum\limits_{m = 1}^{{N_r}} {{{\left( { - 1} \right)}^{m + 1}}{{\left( {\frac{\rho }{{4{{\sin }^2}\left( \vartheta  \right)}}{\lambda _{{n_r}}}} \right)}^{ - m}}}  +
\mathcal{O}\left( {{{\left( { - 1} \right)}^{{N_r} + 1}}{{\left( {\frac{\rho }{{4{{\sin }^2}\left( \vartheta  \right)}}{\lambda _{{n_r}}}} \right)}^{ - ({N_r} + 1)}}} \right)
\end{align}

We note that the following holds true:
\begin{align}
\frac{1}{\pi }\int\limits_0^{\pi /2} {{{\left( {\frac{1}{{4{{\sin }^2}\left( \vartheta  \right)}}} \right)}^{ - m}}d\vartheta }  = \frac{1}{2}\left( {\begin{array}{*{20}{c}}
{2m}\\
m
\end{array}} \right) + \sum\limits_{k = 0}^{m - 1} {{{\left( { - 1} \right)}^{m - k}}2\left( {\begin{array}{*{20}{c}}
{2n}\\
k
\end{array}} \right)\frac{{\sin \left( {\pi \left( {m - k} \right)} \right)}}{{2\pi \left( {m - k} \right)}}}
\end{align}
where we have used the following identity \cite[p. 31, eq. (1320,1)]{jeffrey2007table}:
\begin{align}
{\sin ^{2m}}\left( \vartheta  \right) = \frac{1}{{{2^{2m}}}}\left\{ {\sum\limits_{k = 0}^{m - 1} {{{\left( { - 1} \right)}^{m - k}}2\left( {\begin{array}{*{20}{c}}
{2m}\\
k
\end{array}} \right)\cos \left( {2\left( {m - k} \right)\vartheta } \right) + } \left( {\begin{array}{*{20}{c}}
{2m}\\
m
\end{array}} \right)} \right\}
\end{align}

Let us define:
\begin{align}
{\alpha _m} = \left( {\frac{1}{\pi }\int_0^{\pi /2} {{{\left( {\frac{1}{{4{{\sin }^2}\left( \vartheta  \right)}}} \right)}^{ - m}}d\vartheta } } \right)
\end{align}

Then, we have the following:
\begin{align}
\APEP &\approx {\Exx_{\theta _k^t,\phi _k^t,\theta _k^r,\phi _k^r}}\left\{ {\sum\limits_{{n_r} = 1}^{{N_r}} {{A_{{n_r}}}\left[ {\sum\limits_{m = 1}^{{N_r}} {{{\left( { - 1} \right)}^{m + 1}}{{\left( {\rho {\lambda _{{n_r}}}} \right)}^{ - m}}{\alpha _m}} } \right]} } \right\} \notag\\
 &= {\Exx_{\theta _k^t,\phi _k^t,\theta _k^r,\phi _k^r}}\left\{ {\sum\limits_{{n_r} = 1}^{{N_r}} {{A_{{n_r}}}\left[ {{{\left( { - 1} \right)}^{{N_r} + 1}}{{\left( {\rho {\lambda _{{n_r}}}} \right)}^{ - {N_r}}}{\alpha _{{N_r}}} + \sum\limits_{m = 1}^{{N_r} - 1} {{{\left( { - 1} \right)}^{m + 1}}{{\left( {\rho {\lambda _{{n_r}}}} \right)}^{ - m}}{\alpha _m}} } \right]} } \right\} \notag\\
 &= {\Exx_{\theta _k^t,\phi _k^t,\theta _k^r,\phi _k^r}}\left\{ {\sum\limits_{{n_r} = 1}^{{N_r}} {{A_{{n_r}}}{{\left( { - 1} \right)}^{{N_r} + 1}}{{\left( {\rho {\lambda _{{n_r}}}} \right)}^{ - {N_r}}}{\alpha _{{N_r}}}}  + \sum\limits_{n = 1}^{{N_r}} {{A_n}\left[ {\sum\limits_{m = 1}^{{N_r} - 1} {{{\left( { - 1} \right)}^{m + 1}}{{\left( {\rho {\lambda _{{n_r}}}} \right)}^{ - m}}{\alpha _m}} } \right]} } \right\}
\end{align}

By induction, the following can be proved:
\begin{align}
\sum\limits_{{n_r} = 1}^{{N_r}} {{A_{{n_r}}}\left[ {\sum\limits_{m = 1}^{{N_r} - 1} {{{\left( { - 1} \right)}^{m + 1}}{{\left( {\rho {\lambda _{{n_r}}}} \right)}^{ - m}}{\alpha _m}} } \right]}  = 0
\end{align}
and
\begin{align}
\sum\limits_{{n_r} = 1}^{{N_r}} {{A_{{n_r}}}{{\left( { - 1} \right)}^{{N_r} + 1}}{{\left( {\rho {\lambda _{{n_r}}}} \right)}^{ - {N_r}}}{\alpha _{{N_r}}}}  = \frac{{{\alpha _{{N_r}}}}}{{{\rho ^{{N_r}}}}}\frac{1}{{\prod\limits_{n = 1}^{{N_r}} {{\lambda _n}} }}
\end{align}

\textit{Step 2}: Let $X = \prod\limits_{{n_r} = 1}^{{N_r}} {{\lambda _{{n_r}}}}$. By using steps similar to those of Lemma 2, we obtain the following:
\begin{align}
\det \left( {{\R_{\v}}} \right) = \frac{1}{{{K^{{N_r}}}}}\left( {\sum\limits_{{k_1} = 1}^K { \ldots \sum\limits_{{k_{{N_r}}} = 1}^K {\left( {\prod\limits_{i = 1}^{{N_r}} {{\chi}\left( {{\theta _{{k_i}}}^t,\phi _{{k_i}}^t} \right)F\left( {\theta _{{k_1}}^r, \ldots ,\theta _{{k_{{N_r}}}}^r,\phi _{{k_1}}^r, \ldots ,\phi _{{k_{{N_r}}}}^r} \right)} } \right)} } } \right)
\end{align}
where $F\left( {\theta _{{k_1}}^r, \ldots ,\theta _{{k_{{N_r}}}}^r,\phi _{{k_1}}^r, \ldots ,\phi _{{k_{{N_r}}}}^r} \right)$.

Step 3: The APEP is then the following:
\begin{align}
&\Exx_X\left\{ {\exp \left( {Xz} \right)} \right\}\notag\\
&={\Exx_{{\theta ^t},{\phi ^t},{\theta ^r},{\phi ^r}}}\left\{ {\exp \left( {\frac{z}{{{K^{{N_r}}}}}\left( {\sum\limits_{{k_1} = 1}^K { \ldots \sum\limits_{{k_{{N_r}}} = 1}^K {\left[ {\left( {\prod\limits_{i = 1}^{{N_r}} {{\chi}\left( {{\theta _{{k_i}}}^t,\phi _{{k_i}}^t} \right)} } \right)F\left( {\theta _{{k_1}}^r, \ldots ,\theta _{{k_{{N_r}}}}^r,\phi _{{k_1}}^r, \ldots ,\phi _{{k_{{N_r}}}}^r} \right)} \right]} } } \right)} \right)} \right\} \notag\\
&={\Exx_{{\theta ^t},{\phi ^t},{\theta ^r},{\phi ^r}}}\left\{ {\prod\limits_{{k_1} = 1}^K  \ldots  \prod\limits_{{k_{{N_r}}} = 1}^K {\exp \left( {\frac{z}{{{K^{{N_r}}}}}\left[ {\prod\limits_{i = 1}^{{N_r}} {{\chi}\left( {{\theta _{{k_i}}}^t,\phi _{{k_i}}^t} \right)} } \right]F\left( {\theta _{{k_1}}^r, \ldots ,\theta _{{k_{{N_r}}}}^r,\phi _{{k_1}}^r, \ldots ,\phi _{{k_{{N_r}}}}^r} \right)} \right)} } \right\} \notag\\
&=\prod\limits_{{k_1} = 1}^K  \ldots  \prod\limits_{{k_{{N_r}}} = 1}^K {{\Exx_{{\theta ^t},{\phi ^t},{\theta ^r},{\phi ^r}}}\left\{ {\exp \left( {\frac{z}{{{K^{{N_r}}}}}\left[ {\prod\limits_{i = 1}^{{N_r}} {{\chi}\left( {{\theta _{{k_i}}}^t,\phi _{{k_i}}^t} \right)} } \right]F\left( {\theta _{{k_1}}^r, \ldots ,\theta _{{k_{{N_r}}}}^r,\phi _{{k_1}}^r, \ldots ,\phi _{{k_{{N_r}}}}^r} \right)} \right)} \right\}} \notag\\
&={\left[ {{\Exx_{{\theta ^t},{\phi ^t},{\theta ^r},{\phi ^r}}}\left\{ {\exp \left( {\frac{z}{{{K^{{N_r}}}}}\left[ {\prod\limits_{i = 1}^{{N_r}} {{\chi^i}\left( {{\theta ^{i,t}},{\phi ^{i,t}}} \right)} } \right]F\left( {{\theta ^{1,r}}, \ldots ,{\theta ^{{N_r},r}},{\phi ^{1,r}}, \ldots ,{\phi ^{{N_r},r}}} \right)} \right)} \right\}} \right]^{{N_r}}}
\end{align}

By using the following approximation:
\begin{align}
&\exp \left( {\frac{z}{{{K^{{N_r}}}}}\left[ {\prod\limits_{i = 1}^{{N_r}} {{\chi}\left( {{\theta ^{i,t}},{\phi ^{i,t}}} \right)} } \right]F\left( {{\theta ^{1,r}}, \ldots ,{\theta ^{{N_r},r}},{\phi ^{1,r}}, \ldots ,{\phi ^{{N_r},r}}} \right)} \right) \approx \\
&1 + \frac{z}{{{K^{{N_r}}}}}\left[ {\prod\limits_{i = 1}^{{N_r}} {{\chi}\left( {{\theta ^{i,t}},{\phi ^{i,t}}} \right)} } \right]F\left( {{\theta ^{1,r}}, \ldots ,{\theta ^{{N_r},r}},{\phi ^{1,r}}, \ldots ,{\phi ^{{N_r},r}}} \right)
\end{align}

we have:
\begin{align}
\Exx\left\{ {\exp \left( {Xz} \right)} \right\}&={\left[ {{\Exx_{{\theta ^t},{\phi ^t},{\theta ^r},{\phi ^r}}}\left\{ {1 + \frac{z}{{{K^{{N_r}}}}}\left[ {\prod\limits_{i = 1}^{{N_r}} {{\chi}\left( {{\theta ^{i,t}},{\phi ^{i,t}}} \right)} } \right]F\left( {{\theta ^{1,r}}, \ldots ,{\theta ^{{N_r},r}},{\phi ^{1,r}}, \ldots ,{\phi ^{{N_r},r}}} \right)} \right\}} \right]^{{N_r}}} \notag\\
&={\left[ {1 + \frac{z}{{{K^{{N_r}}}}}{\Exx_{{\theta ^t},{\phi ^t},{\theta ^r},{\phi ^r}}}\left\{ {\left[ {\prod\limits_{i = 1}^{{N_r}} {{\chi}\left( {{\theta ^{i,t}},{\phi ^{i,t}}} \right)} } \right]F\left( {{\theta ^{1,r}}, \ldots ,{\theta ^{{N_r},r}},{\phi ^{1,r}}, \ldots ,{\phi ^{{N_r},r}}} \right)} \right\}} \right]^{{N_r}}} \notag\\
&\approx \exp \left( {z{\Exx_{{\theta ^t},{\phi ^t},{\theta ^r},{\phi ^r}}}\left\{ {\left[ {\prod\limits_{i = 1}^{{N_r}} {{\chi}\left( {{\theta ^{i,t}},{\phi ^{i,t}}} \right)} } \right]F\left( {{\theta ^{1,r}}, \ldots ,{\theta ^{{N_r},r}},{\phi ^{1,r}}, \ldots ,{\phi ^{{N_r},r}}} \right)} \right\}} \right)
\end{align}

Thus, we obtain:
\begin{align}
&\frac{{{\alpha _{{N_r}}}}}{{{\rho ^{{N_r}}}}}\int_0^\infty  {{M_X}\left( { - z} \right)dz}  = \notag\\
&\frac{{{\alpha _{{N_r}}}}}{{{\rho ^{{N_r}}}}}\int_0^\infty  {\exp \left( {z{\Exx_{{\theta ^t},{\phi ^t},{\theta ^r},{\phi ^r}}}\left\{ {\left[ {\prod\limits_{i = 1}^{{N_r}} {{\chi}\left( {{\theta ^{i,t}},{\phi ^{i,t}}} \right)} } \right]F\left( {{\theta ^{1,r}}, \ldots ,{\theta ^{{N_r},r}},{\phi ^{1,r}}, \ldots ,{\phi ^{{N_r},r}}} \right)} \right\}} \right)dz} \notag\\
&= \frac{{{\alpha _{{N_r}}}}}{{{\rho ^{{N_r}}}{\Exx_{{\theta ^t},{\phi ^t},{\theta ^r},{\phi ^r}}}\left\{ {\left[ {\prod\limits_{i = 1}^{{N_r}} {{\chi}\left( {{\theta ^{i,t}},{\phi ^{i,t}}} \right)} } \right]F\left( {{\theta ^{1,r}}, \ldots ,{\theta ^{{N_r},r}},{\phi ^{1,r}}, \ldots ,{\phi ^{{N_r},r}}} \right)} \right\}}} \notag\\
&= \frac{{{\alpha _{{N_r}}}}}{{{\rho ^{{N_r}}}{\Exx_{{\theta ^t},{\phi ^t}}}\left\{ {\left[ {\prod\limits_{i = 1}^{{N_r}} {{\chi}\left( {{\theta ^{i,t}},{\phi ^{i,t}}} \right)} } \right]} \right\}{\Exx_{{\theta ^r},{\phi ^r}}}\left\{ {F\left( {{\theta ^{1,r}}, \ldots ,{\theta ^{{N_r},r}},{\phi ^{1,r}}, \ldots ,{\phi ^{{N_r},r}}} \right)} \right\}}} \notag\\
&= \frac{{{\alpha _{{N_r}}}}}{{{\rho ^{{N_r}}}{{\left[ {{\Exx_{{\theta ^t},{\phi ^t}}}\left\{ {\chi\left( {{\theta ^t},{\phi ^t}} \right)} \right\}} \right]}^{{N_r}}}{\Exx_{{\theta ^r},{\phi ^r}}}\left\{ {F\left( {{\theta ^{1,r}}, \ldots ,{\theta ^{{N_r},r}},{\phi ^{1,r}}, \ldots ,{\phi ^{{N_r},r}}} \right)} \right\}}}
\end{align}

This concludes the proof.

\subsection{Proof of Proposition \ref{pro:APEP2x2}}
\label{subsec:Pro4}
From Theorem \ref{the:Nr}, we can obtain \eqref{eq:APEP2x3}. We need to prove how obtain the explicit form of the function $F(\cdot)$. If $N_r = 3$, the determinant of the covariance matrix $\R$ can be computed as follows:
\begin{align} \label{eq:detR3}
&\det \left( {{\R}} \right) = \notag\\
&\resizebox{\textwidth}{!}{$\Exx\left\{ {\left. {{\nu _1}{\nu ^*}_1} \right\}} \right.\Exx\left\{ {\left. {{\nu _2}{\nu ^*}_2} \right\}} \right.\Exx\left\{ {\left. {{\nu _3}{\nu ^*}_3} \right\}} \right. + \Exx\left\{ {\left. {{\nu _1}{\nu ^*}_2} \right\}} \right.\Exx\left\{ {\left. {{\nu _2}{\nu ^*}_3} \right\}} \right.\Exx\left\{ {\left. {{\nu _3}{\nu ^*}_1} \right\}} \right. + \Exx\left\{ {\left. {{\nu _1}{\nu ^*}_3} \right\}} \right.\Exx\left\{ {\left. {{\nu _2}{\nu ^*}_1} \right\}} \right.\Exx\left\{ {\left. {{\nu _3}{\nu ^*}_2} \right\}} \right.$} \notag\\
&\resizebox{\textwidth}{!}{$- \Exx\left\{ {\left. {{\nu _1}{\nu ^*}_3} \right\}} \right.\Exx\left\{ {\left. {{\nu _2}{\nu ^*}_2} \right\}\Exx\left\{ {\left. {{\nu _3}{\nu ^*}_1} \right\}} \right. - } \right.\Exx\left\{ {\left. {{\nu _1}{\nu ^*}_2} \right\}} \right.\Exx\left\{ {\left. {{\nu _2}{\nu ^*}_1} \right\}} \right.\Exx\left\{ {\left. {{\nu _3}{\nu ^*}_3} \right\}} \right. - \Exx\left\{ {\left. {{\nu _1}{\nu ^*}_1} \right\}} \right.\Exx\left\{ {\left. {{\nu _2}{\nu ^*}_3} \right\}} \right.\Exx\left\{ {\left. {{\nu _3}{\nu ^*}_2} \right\}} \right.$}
\end{align}

We note that:
\begin{align} \label{eq:vuv}
\Exx\left\{ {\left. {{\nu _u}{\nu^{*}_v}} \right\}} \right. = { \frac{1}{K}\sum\limits_{k = 1}^K {{\chi\left( {{\theta ^t_k},{\phi ^t_k}} \right)}} \exp \left( {j\left( {u - v} \right)\left\| \k \right\|{d}\sin \left( {\theta _k^r} \right)\sin \left( {\phi _k^r} \right)} \right)}
\end{align}
for $u, v = 1, \ldots, N_r$. Then from (\ref{eq:detR3}) and (\ref{eq:vuv}), we have:
\begin{align}
\det(\R)={\frac{1}{{{K^3}}}\sum\limits_{{k_1} = 1}^K {\sum\limits_{{k_2} = 1}^K {\sum\limits_{{k_3} = 1}^K {{\chi}\left( {{\theta _{{k_1}}}^t,\phi _{{k_1}}^t} \right)} } } {\chi}\left( {{\theta _{{k_2}}}^t,\phi _{{k_2}}^t} \right){\chi}\left( {{\theta _{{k_3}}}^t,\phi _{{k_3}}^t} \right)F\left( {\theta _{{k_1}}^r,\theta _{{k_2}}^r,\theta _{{k_3}}^r,\phi _{{k_1}}^r,\phi _{{k_2}}^r,\phi _{{k_3}}^r} \right)}
\end{align}
where:
\begin{align}
&F\left( {\theta _{{k_1}}^r,\theta _{{k_2}}^r,\theta _{{k_3}}^r,\phi _{{k_1}}^r,\phi _{{k_2}}^r,\phi _{{k_3}}^r} \right) = \notag\\
&\left[ \begin{array}{l}
1\\
 + \exp \left( { - j\left\| k \right\|{d}\left[ {\sin \left( {\theta _{{k_1}}^r} \right)\sin \left( {\phi _{{k_1}}^r} \right) + \sin \left( {\theta _{{k_2}}^r} \right)\sin \left( {\phi _{{k_2}}^r} \right) - 2\sin \left( {\theta _{{k_3}}^r} \right)\sin \left( {\phi _{{k_3}}^r} \right)} \right]} \right)\\
 + \exp \left( { - j\left\| k \right\|{d}\left[ {2\sin \left( {\theta _{{k_1}}^r} \right)\sin \left( {\phi _{{k_1}}^r} \right) - \sin \left( {\theta _{{k_2}}^r} \right)\sin \left( {\phi _{{k_2}}^r} \right) - \sin \left( {\theta _{{k_3}}^r} \right)\sin \left( {\phi _{{k_3}}^r} \right)} \right]} \right)\\
 - \exp \left( { - j\left\| k \right\|{d}\left[ {2\sin \left( {\theta _{{k_1}}^r} \right)\sin \left( {\phi _{{k_1}}^r} \right) - 2\sin \left( {\theta _{{k_3}}^r} \right)\sin \left( {\phi _{{k_3}}^r} \right)} \right]} \right)\\
 - \exp \left( { - j\left\| k \right\|{d}\left[ {\sin \left( {\theta _{{k_1}}^r} \right)\sin \left( {\phi _{{k_1}}^r} \right) - \sin \left( {\theta _{{k_2}}^r} \right)\sin \left( {\phi _{{k_2}}^r} \right)} \right]} \right)\\
 - \exp \left( { - j\left\| k \right\|{d}\left[ {\sin \left( {\theta _{{k_2}}^r} \right)\sin \left( {\phi _{{k_2}}^r} \right) - \sin \left( {\theta _{{k_3}}^r} \right)\sin \left( {\phi _{{k_3}}^r} \right)} \right]} \right)
\end{array} \right]
\end{align}
which can be further simplified as follows:
\begin{align}
&F\left( {\theta _{{k_1}}^r,\theta _{{k_2}}^r,\theta _{{k_3}}^r,\phi _{{k_1}}^r,\phi _{{k_2}}^r,\phi _{{k_3}}^r} \right) = \notag\\
&\left[ \begin{array}{l}
1\\
 + \cos \left( { - \left\| k \right\|{d}\left[ {\sin \left( {\theta _{{k_1}}^r} \right)\sin \left( {\phi _{{k_1}}^r} \right) + \sin \left( {\theta _{{k_2}}^r} \right)\sin \left( {\phi _{{k_2}}^r} \right) - 2\sin \left( {\theta _{{k_3}}^r} \right)\sin \left( {\phi _{{k_3}}^r} \right)} \right]} \right)\\
 + \cos \left( { - \left\| k \right\|{d}\left[ {2\sin \left( {\theta _{{k_1}}^r} \right)\sin \left( {\phi _{{k_1}}^r} \right) - \sin \left( {\theta _{{k_2}}^r} \right)\sin \left( {\phi _{{k_2}}^r} \right) - \sin \left( {\theta _{{k_3}}^r} \right)\sin \left( {\phi _{{k_3}}^r} \right)} \right]} \right)\\
 - \cos \left( { - \left\| k \right\|{d}\left[ {2\sin \left( {\theta _{{k_1}}^r} \right)\sin \left( {\phi _{{k_1}}^r} \right) - 2\sin \left( {\theta _{{k_3}}^r} \right)\sin \left( {\phi _{{k_3}}^r} \right)} \right]} \right)\\
 - \cos \left( { - \left\| k \right\|{d}\left[ {\sin \left( {\theta _{{k_1}}^r} \right)\sin \left( {\phi _{{k_1}}^r} \right) - \sin \left( {\theta _{{k_2}}^r} \right)\sin \left( {\phi _{{k_2}}^r} \right)} \right]} \right)\\
 - \cos \left( { - \left\| k \right\|{d}\left[ {\sin \left( {\theta _{{k_2}}^r} \right)\sin \left( {\phi _{{k_2}}^r} \right) - \sin \left( {\theta _{{k_3}}^r} \right)\sin \left( {\phi _{{k_3}}^r} \right)} \right]} \right)
\end{array} \right]
\end{align}
where we have used similar observations as for $N_r = 2$.

Now, we need to compute $\Exx_{{\theta ^r},{\phi ^r}}\left( {F\left( {{\theta ^{1,r}},{\theta ^{2,r}},{\theta ^{3,r}},{\phi ^{1,r}},{\phi ^{2,r}},{\phi ^{3,r}}} \right)} \right)$. Let us define the following:
\begin{align}
\begin{array}{l}
a = \left\| k \right\|{d}\sin \left( {\theta _{{k_1}}^r} \right)\sin \left( {\phi _{{k_1}}^r} \right)\\
b = \left\| k \right\|{d}\sin \left( {\theta _{{k_2}}^r} \right)\sin \left( {\phi _{{k_2}}^r} \right)\\
c = \left\| k \right\|{d}\sin \left( {\theta _{{k_3}}^r} \right)\sin \left( {\phi _{{k_3}}^r} \right)
\end{array}
\end{align}

By applying the following trigonometric identities:
\begin{align}
&\resizebox{\textwidth}{!}{$\cos \left( {a + b - 2c} \right) = \cos \left( a \right)\cos \left( b \right)\cos \left( {2c} \right) -\notag \sin \left( a \right)\sin \left( b \right)\cos \left( {2c} \right) +
\sin \left( a \right)\cos \left( b \right)\sin \left( {2c} \right) + \sin \left( b \right)\cos \left( a \right)\sin \left( {2c} \right)$} \\
&\resizebox{\textwidth}{!}{$\cos \left( {c + b - 2a} \right) = \cos \left( c \right)\cos \left( b \right)\cos \left( {2a} \right) - \sin \left( c \right)\sin \left( b \right)\cos \left( {2a} \right) + \sin \left( c \right)\cos \left( b \right)\sin \left( {2a} \right) + \sin \left( b \right)\cos \left( c \right)\sin \left( {2a} \right)$}
\end{align}
and noting that $a, b$ and $c$ are independent, we eventually obtain the desired result after some algebraic manipulations.


\begin{backmatter}

	
	\bibliographystyle{bmc-mathphys} 
	

	

\end{backmatter}
\end{document}

%% file: defs.tex
\DeclareMathOperator{\ABEP}{ABEP}
\DeclareMathOperator{\APEP}{APEP}

\def\y{{\mathbf y}}

\def\e{{\mathbf e}}

\def\w{{\mathbf w}}

\def\a{{\mathbf a}}
\def\k{{\mathbf k}}

\def\A{{\mathbf A}}

\def\n{{\mathbf n}}
\def\R{{\mathbf R}}
\def\I{{\mathbf I}}
\def\v{{\mathbf v}}

\def\H{{\mathbf H}}

\def\0{{\mathbf 0}}

\def\Exx{\mathbb{E}}
\newtheorem{theorem}{Theorem}
\newtheorem{proposition}{Proposition}
\newtheorem{lemma}{Lemma}

%% file: RectAnt_SM.bbl

\begin{thebibliography}{0}
\ifx \bisbn   \undefined \def \bisbn  #1{ISBN #1}\fi
\ifx \binits  \undefined \def \binits#1{#1}\fi
\ifx \bauthor  \undefined \def \bauthor#1{#1}\fi
\ifx \batitle  \undefined \def \batitle#1{#1}\fi
\ifx \bjtitle  \undefined \def \bjtitle#1{#1}\fi
\ifx \bvolume  \undefined \def \bvolume#1{\textbf{#1}}\fi
\ifx \byear  \undefined \def \byear#1{#1}\fi
\ifx \bissue  \undefined \def \bissue#1{#1}\fi
\ifx \bfpage  \undefined \def \bfpage#1{#1}\fi
\ifx \blpage  \undefined \def \blpage #1{#1}\fi
\ifx \burl  \undefined \def \burl#1{\textsf{#1}}\fi
\ifx \doiurl  \undefined \def \doiurl#1{\textsf{#1}}\fi
\ifx \betal  \undefined \def \betal{\textit{et al.}}\fi
\ifx \binstitute  \undefined \def \binstitute#1{#1}\fi
\ifx \binstitutionaled  \undefined \def \binstitutionaled#1{#1}\fi
\ifx \bctitle  \undefined \def \bctitle#1{#1}\fi
\ifx \beditor  \undefined \def \beditor#1{#1}\fi
\ifx \bpublisher  \undefined \def \bpublisher#1{#1}\fi
\ifx \bbtitle  \undefined \def \bbtitle#1{#1}\fi
\ifx \bedition  \undefined \def \bedition#1{#1}\fi
\ifx \bseriesno  \undefined \def \bseriesno#1{#1}\fi
\ifx \blocation  \undefined \def \blocation#1{#1}\fi
\ifx \bsertitle  \undefined \def \bsertitle#1{#1}\fi
\ifx \bsnm \undefined \def \bsnm#1{#1}\fi
\ifx \bsuffix \undefined \def \bsuffix#1{#1}\fi
\ifx \bparticle \undefined \def \bparticle#1{#1}\fi
\ifx \barticle \undefined \def \barticle#1{#1}\fi
\ifx \bconfdate \undefined \def \bconfdate #1{#1}\fi
\ifx \botherref \undefined \def \botherref #1{#1}\fi
\ifx \url \undefined \def \url#1{\textsf{#1}}\fi
\ifx \bchapter \undefined \def \bchapter#1{#1}\fi
\ifx \bbook \undefined \def \bbook#1{#1}\fi
\ifx \bcomment \undefined \def \bcomment#1{#1}\fi
\ifx \oauthor \undefined \def \oauthor#1{#1}\fi
\ifx \citeauthoryear \undefined \def \citeauthoryear#1{#1}\fi
\ifx \endbibitem  \undefined \def \endbibitem {}\fi
\ifx \bconflocation  \undefined \def \bconflocation#1{#1}\fi
\ifx \arxivurl  \undefined \def \arxivurl#1{\textsf{#1}}\fi
\csname PreBibitemsHook\endcsname

\end{thebibliography}

\newcommand{\BMCxmlcomment}[1]{}

\BMCxmlcomment{

<refgrp>

</refgrp>
} 



\begin{thebibliography}{13}
\ifx \bisbn   \undefined \def \bisbn  #1{ISBN #1}\fi
\ifx \binits  \undefined \def \binits#1{#1}\fi
\ifx \bauthor  \undefined \def \bauthor#1{#1}\fi
\ifx \batitle  \undefined \def \batitle#1{#1}\fi
\ifx \bjtitle  \undefined \def \bjtitle#1{#1}\fi
\ifx \bvolume  \undefined \def \bvolume#1{\textbf{#1}}\fi
\ifx \byear  \undefined \def \byear#1{#1}\fi
\ifx \bissue  \undefined \def \bissue#1{#1}\fi
\ifx \bfpage  \undefined \def \bfpage#1{#1}\fi
\ifx \blpage  \undefined \def \blpage #1{#1}\fi
\ifx \burl  \undefined \def \burl#1{\textsf{#1}}\fi
\ifx \doiurl  \undefined \def \doiurl#1{\textsf{#1}}\fi
\ifx \betal  \undefined \def \betal{\textit{et al.}}\fi
\ifx \binstitute  \undefined \def \binstitute#1{#1}\fi
\ifx \binstitutionaled  \undefined \def \binstitutionaled#1{#1}\fi
\ifx \bctitle  \undefined \def \bctitle#1{#1}\fi
\ifx \beditor  \undefined \def \beditor#1{#1}\fi
\ifx \bpublisher  \undefined \def \bpublisher#1{#1}\fi
\ifx \bbtitle  \undefined \def \bbtitle#1{#1}\fi
\ifx \bedition  \undefined \def \bedition#1{#1}\fi
\ifx \bseriesno  \undefined \def \bseriesno#1{#1}\fi
\ifx \blocation  \undefined \def \blocation#1{#1}\fi
\ifx \bsertitle  \undefined \def \bsertitle#1{#1}\fi
\ifx \bsnm \undefined \def \bsnm#1{#1}\fi
\ifx \bsuffix \undefined \def \bsuffix#1{#1}\fi
\ifx \bparticle \undefined \def \bparticle#1{#1}\fi
\ifx \barticle \undefined \def \barticle#1{#1}\fi
\ifx \bconfdate \undefined \def \bconfdate #1{#1}\fi
\ifx \botherref \undefined \def \botherref #1{#1}\fi
\ifx \url \undefined \def \url#1{\textsf{#1}}\fi
\ifx \bchapter \undefined \def \bchapter#1{#1}\fi
\ifx \bbook \undefined \def \bbook#1{#1}\fi
\ifx \bcomment \undefined \def \bcomment#1{#1}\fi
\ifx \oauthor \undefined \def \oauthor#1{#1}\fi
\ifx \citeauthoryear \undefined \def \citeauthoryear#1{#1}\fi
\ifx \endbibitem  \undefined \def \endbibitem {}\fi
\ifx \bconflocation  \undefined \def \bconflocation#1{#1}\fi
\ifx \arxivurl  \undefined \def \arxivurl#1{\textsf{#1}}\fi
\csname PreBibitemsHook\endcsname

\bibitem{koon}
\begin{barticle}
\bauthor{\bsnm{Koonin}, \binits{E.V.}},
\bauthor{\bsnm{Altschul}, \binits{S.F.}},
\bauthor{\bsnm{Bork}, \binits{P.}}:
\batitle{Brca1 protein products: functional motifs}.
\bjtitle{Nat Genet}
\bvolume{13},
\bfpage{266}--\blpage{267}
(\byear{1996})
\end{barticle}
\endbibitem

\bibitem{khar}
\begin{botherref}
\oauthor{\bsnm{Kharitonov}, \binits{S.A.}},
\oauthor{\bsnm{Barnes}, \binits{P.J.}}:
Clinical Aspects of Exhaled Nitric Oxide.
in press
\end{botherref}
\endbibitem

\bibitem{zvai}
\begin{barticle}
\bauthor{\bsnm{Zvaifler}, \binits{N.J.}},
\bauthor{\bsnm{Burger}, \binits{J.A.}},
\bauthor{\bsnm{Marinova-Mutafchieva}, \binits{L.}},
\bauthor{\bsnm{Taylor}, \binits{P.}},
\bauthor{\bsnm{Maini}, \binits{R.N.}}:
\batitle{Mesenchymal cells, stromal derived factor-1 and rheumatoid arthritis
  [abstract]}.
\bjtitle{Arthritis Rheum}
\bvolume{42},
\bfpage{250}
(\byear{1999})
\end{barticle}
\endbibitem

\bibitem{xjon}
\begin{bchapter}
\bauthor{\bsnm{Jones}, \binits{X.}}:
\bctitle{Zeolites and synthetic mechanisms}.
In: \beditor{\bsnm{Smith}, \binits{Y.}} (ed.)
\bbtitle{Proceedings of the First National Conference on Porous Sieves: 27-30
  June 1996; Baltimore},
pp. \bfpage{16}--\blpage{27}
(\byear{1996}).
\bcomment{Stoneham: Butterworth-Heinemann}
\end{bchapter}
\endbibitem

\bibitem{marg}
\begin{bbook}
\bauthor{\bsnm{Margulis}, \binits{L.}}:
\bbtitle{Origin of Eukaryotic Cells}.
\bpublisher{Yale University Press},
\blocation{New Haven}
(\byear{1970})
\end{bbook}
\endbibitem

\bibitem{oreg}
\begin{barticle}
\bauthor{\bsnm{Orengo}, \binits{C.A.}},
\bauthor{\bsnm{Bray}, \binits{J.E.}},
\bauthor{\bsnm{Hubbard}, \binits{T.}},
\bauthor{\bsnm{LoConte}, \binits{L.}},
\bauthor{\bsnm{Sillitoe}, \binits{I.}}:
\batitle{Analysis and assessment of ab initio three-dimensional prediction,
  secondary structure, and contacts prediction}.
\bjtitle{Proteins}
\bvolume{Suppl 3},
\bfpage{149}--\blpage{170}
(\byear{1999})
\end{barticle}
\endbibitem

\bibitem{schn}
\begin{bchapter}
\bauthor{\bsnm{Schnepf}, \binits{E.}}:
\bctitle{From prey via endosymbiont to plastids: comparative studies in
  dinoflagellates}.
In: \beditor{\bsnm{Lewin}, \binits{R.A.}} (ed.)
\bbtitle{Origins of Plastids}
vol. \bseriesno{2},
\bedition{2nd} edn.,
pp. \bfpage{53}--\blpage{76}.
\bpublisher{Chapman and Hall},
\blocation{New York}
(\byear{1993})
\end{bchapter}
\endbibitem

\bibitem{pond}
\begin{botherref}
Innovative Oncology
\end{botherref}
\endbibitem

\bibitem{smith}
\begin{bbook}
\beditor{\bsnm{Smith}, \binits{Y.}} (ed.):
\bbtitle{Proceedings of the First National Conference on Porous Sieves: 27-30
  June 1996; Baltimore}.
\bpublisher{Butterworth-Heinemann},
\blocation{Stoneham}
(\byear{1996})
\end{bbook}
\endbibitem

\bibitem{hunn}
\begin{bchapter}
\bauthor{\bsnm{Hunninghake}, \binits{G.W.}},
\bauthor{\bsnm{Gadek}, \binits{J.E.}}:
\bctitle{The alveloar macrophage}.
In: \beditor{\bsnm{Harris}, \binits{T.J.R.}} (ed.)
\bbtitle{Cultured Human Cells and Tissues},
pp. \bfpage{54}--\blpage{56}.
\bpublisher{Academic Press},
\blocation{New York}
(\byear{1995}).
\bcomment{Stoner G (Series Editor): Methods and Perspectives in Cell Biology,
  vol 1}
\end{bchapter}
\endbibitem

\bibitem{advi}
\begin{botherref}
Advisory Committee on Genetic Modification:
Annual Report.
London
(1999).
Advisory Committee on Genetic Modification
\end{botherref}
\endbibitem

\bibitem{koha}
\begin{botherref}
\oauthor{\bsnm{Kohavi}, \binits{R.}}:
Wrappers for performance enhancement and obvious decision graphs.
PhD thesis,
Stanford University, Computer Science Department
(1995)
\end{botherref}
\endbibitem

\bibitem{mouse}
\begin{botherref}
The Mouse Tumor Biology Database.
\url{http://tumor.informatics.jax.org/cancer\_links.html}
\end{botherref}
\endbibitem

\end{thebibliography}

\newcommand{\BMCxmlcomment}[1]{}

\BMCxmlcomment{

<refgrp>

<bibl id="B1">
  <title><p>BRCA1 protein products: functional motifs</p></title>
  <aug>
    <au><snm>Koonin</snm><fnm>E V</fnm></au>
    <au><snm>Altschul</snm><fnm>S F</fnm></au>
    <au><snm>Bork</snm><fnm>P</fnm></au>
  </aug>
  <source>Nat Genet</source>
  <pubdate>1996</pubdate>
  <volume>13</volume>
  <fpage>266</fpage>
  <lpage>267</lpage>
</bibl>

<bibl id="B2">
  <title><p>Clinical aspects of exhaled nitric oxide</p></title>
  <aug>
    <au><snm>Kharitonov</snm><fnm>S A</fnm></au>
    <au><snm>Barnes</snm><fnm>P J</fnm></au>
  </aug>
  <source>Eur Respir J</source>
  <inpress />
</bibl>

<bibl id="B3">
  <title><p>Mesenchymal cells, stromal derived factor-1 and rheumatoid
  arthritis [abstract]</p></title>
  <aug>
    <au><snm>Zvaifler</snm><fnm>N J</fnm></au>
    <au><snm>Burger</snm><fnm>J A</fnm></au>
    <au><snm>Marinova Mutafchieva</snm><fnm>L</fnm></au>
    <au><snm>Taylor</snm><fnm>P</fnm></au>
    <au><snm>Maini</snm><fnm>R N</fnm></au>
  </aug>
  <source>Arthritis Rheum</source>
  <pubdate>1999</pubdate>
  <volume>42</volume>
  <fpage>s250</fpage>
</bibl>

<bibl id="B4">
  <title><p>Zeolites and synthetic mechanisms</p></title>
  <aug>
    <au><snm>Jones</snm><fnm>X</fnm></au>
  </aug>
  <source>Proceedings of the First National Conference on Porous Sieves: 27-30
  June 1996; Baltimore</source>
  <editor>Y Smith</editor>
  <pubdate>1996</pubdate>
  <fpage>16</fpage>
  <lpage>27</lpage>
</bibl>

<bibl id="B5">
  <title><p>Origin of Eukaryotic Cells</p></title>
  <aug>
    <au><snm>Margulis</snm><fnm>L</fnm></au>
  </aug>
  <publisher>New Haven: Yale University Press</publisher>
  <pubdate>1970</pubdate>
</bibl>

<bibl id="B6">
  <title><p>Analysis and assessment of ab initio three-dimensional prediction,
  secondary structure, and contacts prediction</p></title>
  <aug>
    <au><snm>Orengo</snm><fnm>C A</fnm></au>
    <au><snm>Bray</snm><fnm>J E</fnm></au>
    <au><snm>Hubbard</snm><fnm>T</fnm></au>
    <au><snm>LoConte</snm><fnm>L</fnm></au>
    <au><snm>Sillitoe</snm><fnm>I</fnm></au>
  </aug>
  <source>Proteins</source>
  <pubdate>1999</pubdate>
  <volume>Suppl 3</volume>
  <fpage>149</fpage>
  <lpage>170</lpage>
</bibl>

<bibl id="B7">
  <title><p>From prey via endosymbiont to plastids: comparative studies in
  dinoflagellates</p></title>
  <aug>
    <au><snm>Schnepf</snm><fnm>E</fnm></au>
  </aug>
  <source>Origins of Plastids</source>
  <publisher>New York: Chapman and Hall</publisher>
  <editor>R A Lewin</editor>
  <edition>2</edition>
  <pubdate>1993</pubdate>
  <volume>2</volume>
  <fpage>53</fpage>
  <lpage>76</lpage>
</bibl>

<bibl id="B8">
  <title><p>Innovative oncology</p></title>
  <source>Breast Cancer Res</source>
  <editor>B Ponder and S Johnston and L Chodosh</editor>
  <pubdate>1998</pubdate>
  <volume>10</volume>
  <fpage>1</fpage>
  <lpage>72</lpage>
</bibl>

<bibl id="B9">
  <title><p>Proceedings of the First National Conference on Porous Sieves:
  27-30 June 1996; Baltimore</p></title>
  <publisher>Stoneham: Butterworth-Heinemann</publisher>
  <editor>Y Smith</editor>
  <pubdate>1996</pubdate>
</bibl>

<bibl id="B10">
  <title><p>The alveloar macrophage</p></title>
  <aug>
    <au><snm>Hunninghake</snm><fnm>G W</fnm></au>
    <au><snm>Gadek</snm><fnm>J E</fnm></au>
  </aug>
  <source>Cultured Human Cells and Tissues</source>
  <publisher>New York: Academic Press</publisher>
  <editor>T J R Harris</editor>
  <pubdate>1995</pubdate>
  <fpage>54</fpage>
  <lpage>56</lpage>
  <note>Stoner G (Series Editor): Methods and Perspectives in Cell Biology, vol
  1</note>
</bibl>

<bibl id="B11">
  <title><p>Annual Report</p></title>
  <aug><au><cnm>Advisory Committee on Genetic Modification</cnm></au></aug>
  <publisher>London</publisher>
  <pubdate>1999</pubdate>
</bibl>

<bibl id="B12">
  <title><p>Wrappers for performance enhancement and obvious decision
  graphs</p></title>
  <aug>
    <au><snm>Kohavi</snm><fnm>R</fnm></au>
  </aug>
  <source>PhD thesis</source>
  <publisher>Stanford University, Computer Science Department</publisher>
  <pubdate>1995</pubdate>
</bibl>

<bibl id="B13">
  <title><p>The Mouse Tumor Biology Database</p></title>
  <url>http://tumor.informatics.jax.org/cancer\_links.html</url>
</bibl>

</refgrp>
} 


\begin{thebibliography}{99}

%
\bibitem{MDR_Magazine} M. Di Renzo, H. Haas, and P. M. Grant,
              ``Spatial modulation for multiple-antenna wireless systems: A survey,''
             \emph{IEEE Commun. Mag.},
             vol. 49, no. 12, pp. 182-191, Dec. 2011.
%
\bibitem{Younis_TCOM2013} A. Younis, S. Sinanovic, M. Di Renzo, R. Y. Mesleh, and H. Haas,
              ``Generalised sphere decoding for spatial modulation,''
             \emph{IEEE Trans. Commun.},
             vol. 61, no. 7, pp. 2805-2815, July 2013.
%
\bibitem{Athanasios_VTC2013} A. Stavridis, S. Sinanovic, M. Di Renzo, and H. Haas,
              ``Energy evaluation of spatial modulation at a multi-antenna base station,''
             \emph{IEEE Veh. Technol. Conf. - Fall},
             pp. 1-5, Sep. 2013.
%
\bibitem{OrangeAccess} D.-T. Phan-Huy \textit{et al.},
              ``Single-carrier spatial modulation for the Internet of Things: Design and performance evaluation by using real compact and reconfigurable antennas,''
             submitted, 2018. [Online]. Available: https://arxiv.org/abs/1812.07514.
%
\bibitem{MDR_HWU} P. Patcharamaneepakorn, S. Wu, C.-X. Wang, E.-H. M. Aggoune, M. M. Alwakeel, X. Ge, and M. Di Renzo,
              ``Spectral, energy, and economic efficiency of 5G multicell massive MIMO systems with generalized spatial modulation,''
             \emph{IEEE Trans. Veh. Technol.},
             vol. 65, no. 12, pp. 9715-9731, Dec. 2016.
%
%
%
%
\bibitem{MDR_ProcIEEE} M. Di Renzo, H. Haas, A. Ghrayeb, S. Sugiura, and L. Hanzo,
              ``Spatial modulation for generalized MIMO: Challenges, opportunities and implementation,''
             \emph{Proc. of the IEEE}
             vol. 102, no. 1, pp. 56--103, Jan. 2014.
%
\bibitem{MDR_CST} P. Yang, M. Di Renzo, Y. Xiao, S. Li, and L. Hanzo,
              ``Design guidelines for spatial modulation,''
             \emph{IEEE Commun. Surveys Tuts.},
             vol. 16, no. 1, pp. 6-26, 1st quarter 2015.
%
\bibitem{MDR_SingleCarrier} P. Yang, Y. Xiao, Y. L. Guan, K. V. S. Hari, A. Chockalingam, S. Sugiura, H. Haas, M. Di Renzo, C. Masouros, Z. Liu, L. Xiao, S. Li, and L. Hanzo,
              ``Single-carrier SM-MIMO: A promising design for broadband large-scale antenna systems,''
             \emph{IEEE Commun. Surveys Tuts.},
             vol. 18, no. 3, pp. 1687-1716, 3rd quarter 2016.
%
\bibitem{MDR_Wiley} M. Di Renzo, H. Haas, A. Ghrayeb, S. Sugiura, and L. Hanzo,
              ``Spatial modulation for multiple-antenna communication,''
             \emph{Wiley Encyclopedia of Electrical and Electronics Engineering},
             Nov. 2016.
%
\bibitem{MDR_IM} E. Basar, M. Wen, R. Mesleh, M. Di Renzo, Y. Xiao, and H. Haas,
              ``Index modulation techniques for next-generation wireless networks,''
             \emph{IEEE Access},
             vol. 5, pp. 16693-16746, 2017.
%
%
%
\bibitem{YounisPractical} A. Younis, W. Thompson, M. Di Renzo, C.-X. Wang, M. A. Beach, H. Haas, and P. M. Grant,
             ``Performance of spatial modulation using measured real-world channels,''
             \emph{IEEE Veh. Technol. Conf. - Fall},
             pp. 1-5, Sep. 2013.
%
\bibitem{SerafimovskiPractical} N. Serafimovski, A. Younis, R. Mesleh, P. Chambers, M. Di Renzo, C.-X. Wang, P. M. Grant, M. A. Beach, and H. Haas,
             ``Practical implementation of spatial modulation,''
             \emph{IEEE Trans. Veh. Technol.},
             vol. 62, no. 9, pp. 4511-4523, Sep. 2013.
%
\bibitem{Liu_1} P. Liu, M. Di Renzo, and A. Springer,
                    ``Line-of-sight spatial modulation for indoor mmWave communication at 60 GHz,''
                    \emph{IEEE Trans. Wireless Commun.},
             		vol. 15, no. 11, pp. 7373-7389, Nov. 2016.
%
\bibitem{Liu_2} P. Liu, J. Blumenstein, N. S. Perovic, Ma. Di Renzo, and A. Springer,
                    ``Performance of generalized spatial modulation MIMO over measured 60 GHz indoor channels,''
                    \emph{IEEE Trans. Commun.},
             		vol. 66, no. 1, pp. 133-148, Jan. 2018.
%
%
%
%
\bibitem{MDR_TVTMar2012} M. Di Renzo and H. Haas,
              ``Bit error probability of SM-MIMO over generalized fading channels,''
             \emph{IEEE Trans. Veh. Technol.},
             vol. 61, no. 3, pp. 1124-1144, Mar. 2012.
%
\bibitem{MDR_TVT2013} M. Di Renzo and H. Haas,
              ``On transmit-diversity for spatial modulation MIMO: Impact of spatial-constellation diagram and shaping filters at the transmitter,''
             \emph{IEEE Trans. Veh. Technol.},
             vol. 62, no. 6, pp. 2507-2531, July 2013.
%
\bibitem{MDRVietnam_TCOM2014} M.-T. Le, V.-D. Ngo, H.-A. Mai, X.-N. Tran, and M. Di Renzo,
             ``Spatially modulated orthogonal space-time block codes with nonvanishing determinants,''
             \emph{IEEE Trans. Commun.},
             vol. 62, no. 1, pp. 85-99, Jan. 2014.
%
\bibitem{MDR_TVT2016} D. A. Basnayaka, M. Di Renzo, and H. Haas,
              ``Massive but few active MIMO,''
             \emph{IEEE Trans. Veh. Technol.},
             vol. 65, no. 9, pp. 6861-6877, Sep. 2016.
%
\bibitem{MDR_TCOM2018} N. S. Perovic, P. Liu, J. Blumenstein, M. Di Renzo, and A. Springer,
              ``Optimization of the cut-off rate of generalized spatial modulation with transmit precoding,''
             \emph{IEEE Trans. Commun.},
             vol. 66, no. 10, pp. 4578-4595, Oct. 2018.
%
%
%
%
\bibitem{MDR_Milcom2017} M. Di Renzo,
              ``Spatial modulation based on reconfigurable antennas - A new air interface for the IoT,''
             \emph{IEEE Military Communications Conference},
             pp. 1-6, Oct. 2017.
%
%
%
%
\bibitem{NKDA2015} Q. U. A. Nadeem, A. Kammoun, M. Debbahn and M. S. Alouini,
             ``A generalized spatial correlation model for 3D MIMO channels based on the Fourier coefficients of power spectrums,''
             \emph{IEEE Trans. Sig. Proc.},
             vol. 63, no. 14, pp. 3671-3686, Jul. 2015.
%
\bibitem{ZPPLC2014} J. Zhang, C. Pan, F. Pei, G. Liu, and X. Cheng,
             ``Three-dimensional fading channel models: A survey of elevation angle research,''
             \emph{IEEE Commun. Mag.},
             vol. 52, no. 6, pp. 218-226, Jun. 2014.
%
\bibitem{Pollock2003} T. S. Pollock, T. D. Abhayapala, and R. A. Kennedy,
             ``Three-dimensional fading channel models: A survey of elevation angle research,''
             \emph{Telecommunications Systems},
             vol. 24, no. 2, pp. 415-436, 2003.
%
\bibitem{MMGSC2013} M. Di Renzo, C. Merola, A. Guidotti, F. Santucci, and G. E. Corazza,
             ``Error performance of multi-antenna receivers in a Poisson field of interferers: A stochastic geometry approach,''
             \emph{IEEE Trans. Commun.},
             vol. 61, no. 5, pp. 2025-2047, May 2013.
%
\bibitem{simon2005digital} M. K. Simon and M.-S. Alouini,
             \emph{Digital Communication over Fading Channels},
             John Wiley \& Sons, 2005.
%
%
\bibitem{jeffrey2007table} A. Jeffrey and D. Zwillinger,
             \emph{Table of Integrals, Series, and Products},
             Elsevier, 2007.             
%
\bibitem{Turin1960} G. L. Turin,
             ``The characteristic function of Hermitian quadratic forms in complex normal variables,''
             \emph{Biometrika},
             vol. 67, no. 1/2, pp. 199–201, 1960.
%
%
%
\end{thebibliography}
